\documentclass[12pt]{article}

\pdfoutput=1
\usepackage[utf8]{inputenc}
\usepackage{dsfont}
\usepackage{diagbox}
\usepackage{amsfonts}
\usepackage{mathtools}
\allowdisplaybreaks[4]        
\usepackage{amssymb}
\usepackage{euscript}     
\usepackage{braket}
\usepackage{starfont}
\usepackage{color,soul}         
\usepackage{tensor}        
\usepackage{amsthm}
\usepackage{graphicx}
\usepackage{slashed}
\usepackage{leftidx}
\usepackage{subfigure}
\usepackage{bbm}
\definecolor{outerspace}{rgb}{0.25, 0.29, 0.3}
\definecolor{scarlet}{rgb}{1.0, 0.13, 0.0}
\usepackage[header,title,page,titletoc]{appendix}  
\definecolor{princetonorange}{rgb}{1.0, 0.56, 0.0}
\definecolor{WildStrawberry}{rgb}{1.0, 0.26, 0.64}
\definecolor{rossocorsa}{rgb}{0.83, 0.0, 0.0}
\definecolor{navyblue}{rgb}{0.0, 0.0, 0.5}
\usepackage[numbers,sort&compress]{natbib}  
\usepackage{float}
\usepackage[paper=letterpaper,margin=1in]{geometry}
\usepackage{physics}

\newcommand{\bea}{\begin{eqnarray}}
\newcommand{\eea}{\end{eqnarray}}
\newcommand{\be}{\begin{equation}}
\newcommand{\ee}{\end{equation}}

\usepackage{hyperref}
\hypersetup{
    colorlinks,
    citecolor=rossocorsa,
    filecolor=navyblue,
    linkcolor=navyblue,
    urlcolor=navyblue
}

\begin{document} 

\begin{titlepage}

\begin{center}

\phantom{ }
\vspace{3cm}

{\bf \Large{ Standard translation twists and an operator-bounded energy inequality}}
\vskip 0.5cm
Horacio Casini${}^{\text{\Kronos}}$ and Leandro Martinek${}^{\text{\Zeus}}$
\vskip 0.05in
\textit{Instituto Balseiro, Centro At\'omico Bariloche}
\vskip -.4cm
\textit{ 8400-S.C. de Bariloche, R\'io Negro, Argentina}

\begin{abstract}
Twist operators implement symmetries in bounder regions of the space.  Standard twists are a special class of twists constructed using modular tools. The twists corresponding to translations have interesting special properties. They can move continuously an operator from a region to a disjoint one without ever passing through the gap separating the two. In addition, they have generators satisfying the spectrum condition.  We compute explicitly these twists for the two dimensional chiral fermion field. The twist generator gives place to a new type of energy inequality where the smeared energy density is bounded below by an operator.  
\end{abstract}
\end{center}

\small{\vspace{5cm}\noindent${}^{\text{\text{\Kronos}}}$casini@cab.cnea.gov.ar\\
${}^{\text{\Zeus}}$leandro.martinek@ib.edu.ar}

\end{titlepage}
\setcounter{tocdepth}{2}
{\parskip = .2\baselineskip \tableofcontents}

\section{Introduction}
\label{sec:intro}

A twist is an operator that implements a symmetry in a region of space but does nothing outside. Twists implementing any symmetry, be it internal or spacetime symmetry, can always be constructed in a standard way using modular tools  \cite{doplicher1982local,doplicher1984standard,doplicher1983local,
buchholz1986noether}. It is simple to construct a twist in a spatial region $A$ for a Noether symmetry with Noether current $j^{\mu}$. This is done by exponentiating the local charge
\be
\tau_{A}= e^{i \lambda\, \int d^{d-1}x \, f(x)\,j^0(x)}\,,\label{uno}
\ee
where $f(x)=1$ inside $A$ and $f(x)$ smoothly go to zero outside $A\cup Z$, where $Z$ is a small buffer zone surrounding $A$. This is necessary to have a non-singular operator.  We are also omitting a small smearing in the time direction. Using modular theory it is possible to construct twists implementing locally also discrete symmetries without Noether currents. While the strong form of the Noether theorem (existence of currents for continuous symmetries) is still incomplete \cite{Benedetti:2022zbb,harlow2021symmetries}, this gives a weak form of the Noether theorem that is also valid for discrete symmetries \cite{buchholz1986noether}.   

Using a translation current $j^\mu=a_\nu \,T^{\mu\nu}$, with $T^{\mu\nu}$ the stress tensor, (\ref{uno}) will implement locally translations in the $a_\nu$ direction. Suppose now we have two identical but disjoint balls $A_1$ and $A_2$ in space and want a unitary operator that implements the translation from $A_1$ to $A_2$ on all operators localized on $A_1$, but does nothing outside of $(A_1\cup Z_1)\cup (A_2\cup Z_2)$, for two arbitrarily small buffer zones $Z_1, Z_2$ surrounding the balls. That is, we want a translation operator capable of leaping from one region to the other. Is this possible?  

By integrating the charge density over the two balls, we will never be able to translate operators from $A_1$ to $A_2$. Operators from $A_{1}$ will get crushed into the buffer zone $Z_1$ of $A_1$ for large translations and never reach $A_2$. This Noether twist is a product of two operators localized in $A_1\cup Z_1$ and $A_2\cup Z_2$, which makes such a translation impossible. However, it is the result of a paper by Buchholz, Doplicher and Longo \cite{buchholz1986noether} that such a translation operator generically exists in quantum field theory (QFT). 
 The generator of this translation from $A_1$ to $A_2$ should contain something similar to the Noether charge in $A_1\cup A_2$, such as to implement infinitesimal translations, but something more, connecting the two regions, is needed in the buffer zone around the two balls. This is precisely what the standard twist mentioned above achieves. 

This twist is abstractly constructed using modular theory, and as such it is generally difficult to grasp what it is ``made of''. One of the objectives of this paper is to explicitly construct a translation twist in a simple model, the chiral $d=2$ fermion, where the necessary modular tools are sufficiently known.  

Another interest is that standard twists are constructed by unitary transformations of the global symmetry operator by a universal localizing map. Therefore, the twist and its charge generator have the same spectrum as the global ones \cite{buchholz1986noether}. This implies that standard translation local charges satisfy the spectrum condition of the global momentum operator. Understanding these local charges as a contribution proportional to the stress tensor and additional operators will lead us to operator-bounded energy inequalities, where the suitably smeared energy density in some region of the space is bounded below by an operator localized in the boundary. This differs from the usual energy bounds where the energy density is bounded by a number.  
We find these bounds explicitly for the chiral fermion and compare them with the Fewster-Hollands energy density bounds for a CFT \cite{fewster2005quantum}. We find that none of the bounds is generally stronger than the other and compare the conditions for saturation.    

The paper is organized as follows. In section \ref{BDL} we review the construction of standard twists using modular theory. In section \ref{chiral} we describe the modular data for the chiral fermion. In section \ref{explicit} we display how standard translation twists act in the model, and in particular the twists that jump from one interval to another.   
In section \ref{charges} we show the explicit form of the standard twist charges for this model, and study the operator-bounded energy conditions. Finally, we present the conclusions.

\section{Standard twists}
\label{BDL}

Given two separated regions $A$ and $B$ in QFT and the corresponding operator algebras $\mathcal{A}$ and $\mathcal{B}$, the split property asserts the existence of a type I factors $\mathcal{N}$,
$\mathcal{N}'$, such that $\mathcal{A} \subset \mathcal{N}$, $\mathcal{B} \subset \mathcal{N}'$. Here the prime indicates the commutant of an algebra. A
 factor of type I is the algebra of all bounded operators acting in some Hilbert space ${\cal H}$. An equivalent description of the split property is that the factors 
$\mathcal{N}$ and $\mathcal{N}'$ are the algebras acting on two Hilbert spaces $\mathcal{H}_{\mathcal{N}}$, $\mathcal{H}_{\mathcal{N}'}$ such that the global Hilbert space writes
\begin{equation}
    \mathcal{H} = \mathcal{H}_{\mathcal{N}} \otimes  \mathcal{H}_{\mathcal{N}'}  \,\text{.}
\end{equation}
Therefore the algebra $\mathcal{A}$ acts on the first Hilbert space factor and  $\mathcal{B}$ in the second. The split property has been proved to hold in QFT under very general conditions \cite{buchholz1986causal,buchholz1987universal}. 

The factor $\mathcal{N}$ in the split property is highly non-unique. However, a standard construction follows from modular theory that gives this split from the vacuum state $\ket{\Omega}$ and the two algebras \cite{doplicher1984standard}.
The essential tool is modular conjugation which is defined as follows. Given an algebra ${\cal A}$ and a (cyclic separating\footnote{Cyclic means the algebra acting on the vector generates a dense set of vectors in the Hilbert space. Separating is that no non-trivial element of the algebra can annihilate the state.}) vector  $|\Omega \rangle$ the modular conjugation $J$ is defined through
\bea
S \, a|\Omega \rangle &=& a^\dagger |\Omega \rangle\,,\hspace{.7cm}a \in {\cal A}\,,\\
 S &=& J \Delta^{1/2}\,. 
\eea  
The last equation is the polar decomposition of $S$ into a positive operator $\Delta=e^{-K}$, and an anti-unitary $J$, with $J \,\Delta=\Delta^{-1}\, J$. $K$ is known as the modular Hamiltonian. The modular conjugation $J$ maps the algebra into its commutant,  $J {\cal A} J={\cal A}'$ and satisfies $J^2=1$, $J=J^\dagger$. See \cite{Haag:1992hx} for details on modular operators.  
 
In the present case, the vacuum is cyclic and separating for the algebra $\mathcal{AB} \equiv \mathcal{A} \vee \mathcal{B}$, that is, the algebra generated by the two. Then, it induces the modular conjugation $J_{\mathcal{A} \mathcal{B}}$ corresponding to this algebra.  $J_{\mathcal{A} \mathcal{B}}$ maps the algebra of the two regions into its commutant. 
The standard split factors are constructed as \cite{doplicher1984standard}
\begin{equation}
    \mathcal{N} = \mathcal{A} \vee (J_{\mathcal{A} \mathcal{B}}\, \mathcal{A}\, J_{\mathcal{A} \mathcal{B}}) \ \text{,} \quad \mathcal{N}^{\prime} = \mathcal{B} \vee (J_{\mathcal{A} \mathcal{B}} \,\mathcal{B}\, J_{\mathcal{A} \mathcal{B}}) \ \text{.}
\end{equation} 
These factors obey  $J_{\mathcal{A} \,\mathcal{B}}\, \mathcal{N}\, J_{\mathcal{A} \mathcal{B}} = \mathcal{N}$,
$J_{\mathcal{A} \mathcal{B}}\, \mathcal{N^{\prime}}\, J_{\mathcal{A} \mathcal{B}} = \mathcal{N^{\prime}}$, and are defined by this property. Notice the type I factor ${\cal N}$ is generated by the mutually commuting algebras ${\cal A}$ and $J_{\mathcal{A} \mathcal{B}}\, \mathcal{A}\, J_{\mathcal{A} \mathcal{B}}$. 

For the applications we have in mind $A$ is a causal region and both these algebras are type III$_1$ factors, but together they generate ${\cal N}$. The commuting algebras $J_{\mathcal{A} \mathcal{B}}\, \mathcal{A}\, J_{\mathcal{A} \mathcal{B}}$ and $J_{\mathcal{A} \mathcal{B}}\, \mathcal{B}\, J_{\mathcal{A} \mathcal{B}}$ are composed of operators localized in the buffer zone $Z$ that is the complement of the union of $A$ and $B$.    

Consider the vacuum state acting on ${\cal A}$. There is a unique purification of this state on ${\cal A}$ considered inside the Hilbert space ${\cal H}_{\cal N}$, where this purified state is invariant under $J_{\mathcal{A} \mathcal{B}}$. Call this state  $\ket{\Omega}_{\mathcal{N}}$. In an analogous way $\ket{\Omega}_{\mathcal{N}'}$ is defined as the unique vector that purifies the vacuum acting on ${\cal B}$ inside ${\cal H}_{{\cal N}'}$, and that is invariant under $J_{\mathcal{A} \mathcal{B}}$. The vector   
$\ket{\eta} = \ket{\Omega}_{\mathcal{N}} \otimes \ket{\Omega}_{\mathcal{N}^{\prime}}$ acts as the vacuum on ${\cal A}$ or ${\cal B}$ but erases correlations between the algebras. The vectors
 $\ket{\Omega}_{\mathcal{N}}$ and $\ket{\Omega}_{\mathcal{N}^{\prime}}$
are cyclic inside the corresponding factors ${\cal N}$ and ${\cal N}'$. Consequently $\ket{\eta}$ is cyclic in the full Hilbert space. 

The unitary localization map $W: \mathcal{H} \rightarrow \mathcal{H} \otimes \mathcal{H}$
is defined such that for any $a \in \mathcal{A}$ y $b \in \mathcal{B}$ 
\begin{equation}
    W a b \ket{\eta} = a \ket{\Omega} \otimes b \ket{\Omega}  \,\text{.}
\end{equation}
This completely defines the map because the action of ${\cal A}$ and ${\cal B}$ together is cyclic on both sides. It can be shown that $W$ is an isomorphism. $W^\dagger$ maps the factors in the tensor product $\mathcal{H} \otimes \mathcal{H}$ into the the factors defined by the product ${\cal H}_{\cal N}\otimes{\cal H}_{{\cal N}'}$. This is why it can be used to localize in ${\cal N}$ operators defined globally, which can be thought to act on the first factor in  $\mathcal{H} \otimes \mathcal{H}$. 
It follows from this definition that
\be
    W \,a\, W^{\dagger} = a \otimes 1, \quad W\, b\, W^{\dagger} = 1 \otimes b \ \text{,} \quad a \in \mathcal{A} \ \text{,} \ b \in \mathcal{B} \,,\label{prop}
\ee    
and
   \be 
    W \,J_{\mathcal{A} \mathcal{B}}\, W^{\dagger} = J_{{\cal A}} \otimes J_{{\cal B}} \ \text{.}\label{jotas}
\ee

A standard twist can be readily constructed with these ingredients.  
 Given a global unitary $U$ acting on ${\cal H}$ the localization map provides a local unitary $\tau$ acting non trivially only on the first split factor $\mathcal{N}$ and leaving invariant $\mathcal{N}'$:
\begin{equation}
    \tau = W^{\dagger} \left( U \otimes 1 \right) W\, \text{.}
    \label{eq:deftwist}
\end{equation}
If $U$ is a global symmetry $\tau$ will be a twist operator for this symmetry. In fact, let $a \in \mathcal{A}$, $b \in \mathcal{B}$ and assume
 $U\, a\, U^{\dagger} \in \mathcal{A}$. Using (\ref{prop}) it follows that
\bea
    \tau \ a \ \tau^{\dagger} &=& U \ a \ U^{\dagger} \, \text{,} \label{rr}\\
        \tau \ b \ \tau^{\dagger} &=& b \, \text{.}\label{rr1}
\eea

In this way, if $U$ is a translation, and the translation keeps an operator $a$ in the algebra ${\cal A}$, the action of $\tau$ will be exactly as a translation. This includes the case where the region $A$ is formed by more than one connected component and the translation carries an operator from one component to the other. In addition, $\tau$ will do nothing to operators localized in $B$, no matter the size of the applied translation. 

To understand these twists more explicitly it is convenient to describe the operator content of the theory by local field operators $\psi(x)$. In general, we know the action of the symmetry $U$ on the field operators. This will carry field operators into field operators. It remains to have an understanding of the action of the localization map $W$ on them. We already know how $(\psi(x) \otimes 1)$ transforms for $x\in A$, eq. (\ref{prop}). 
To understand the transformation of $(\psi(x) \otimes 1)$ for $x\in A'$ we can first write this operator as an operator $J_{{\cal A}} \, \phi \, J_{{\cal A}}$, for some $\phi$ inside ${\cal A}$ in the first factor, by acting with the modular conjugation of ${\cal A}$. Call $A'$ to the complementary space-like separated region to $A$, such that operators localized in $A'$ commute with ${\cal A}$.  We get for $x \in A'$
 \bea
    W^{\dagger} \left(\psi(x) \otimes 1\right) W &=& 
            W^{\dagger} \left( J_{\mathcal{A}} \ \phi \ J_{\mathcal{A}} \otimes 1\right) W \nonumber\\
            &=&  W^{\dagger} \left(J_{\mathcal{A}} \otimes J_{\mathcal{B}}\right)
            \left(\phi \otimes 1\right) \left(J_{\mathcal{A}} \otimes J_{\mathcal{B}}\right) W \nonumber\\ 
            &=&  W^{\dagger} \left(J_{\mathcal{A}} \otimes J_{\mathcal{B}}\right) W \ W^{\dagger}
            \left(\phi \otimes 1\right)  W \ W^{\dagger} \left(J_{\mathcal{A}} \otimes J_{\mathcal{B}}\right) W \nonumber\\ &=&  J_{\mathcal{A}  \mathcal{B}} \ \phi \ J_{\mathcal{A}  \mathcal{B}}             
            = J_{\mathcal{A}  \mathcal{B}} \ J_{\mathcal{A}}  \ \psi(x)  \ J_{\mathcal{A}} \ J_{\mathcal{A}  \mathcal{B}} \ \text{.}
        \label{aaa}
        \eea
An analogous calculation gives the transformation for a field on $B'$ in the second factor. So we have that
\bea
   W^\dagger\, (\psi(x) \otimes 1)\,W &=& \psi(x)\,, \hspace{3.5cm} x\in A\label{prop0}\,,\\ 
     W^\dagger\, (1 \otimes \psi(x))\, W &=& \psi(x) \,, \hspace{3.5cm}  x \in B\,,\label{prop1}\\
 W^\dagger\, (\psi(x) \otimes 1)\,W &=& J_{\mathcal{A}  \mathcal{B}} \ J_{\mathcal{A}}  \ \psi(x)  \ J_{\mathcal{A}} \ J_{\mathcal{A}  \mathcal{B}}\,, \hspace{.6cm} x\in A'\,, \label{qq}\\ 
     W^\dagger\, (1 \otimes \psi(x))\, W &=& J_{\mathcal{A}  \mathcal{B}} \ J_{\mathcal{B}}  \ \psi(x)  \ J_{\mathcal{B}} \ J_{\mathcal{A}  \mathcal{B}} \,, \hspace{.7cm}  x \in B'\,.
\eea  
 
For Noether symmetries, these rules already give a more explicit form for the twist. Eq. (\ref{eq:deftwist}) gives the twist generator or local charge, $Q_A$, as the mapping of the global symmetry charge operator $Q$ acting on the first factor. Written in terms of the transformed charge density it is:
\bea
Q_A &=& W^{\dagger} \left( Q \otimes 1 \right) W 
=\int d^{d-1}x\,  W^{\dagger} \left( j^0(x) \otimes 1 \right) W\,  \nonumber \\
&=& \int_A d^{d-1}x \, j^0(x)+ \int_{A'} d^{d-1}x \, J_{\mathcal{A}  \mathcal{B}} \ J_{\mathcal{A}}  \ j^0(x)  \ J_{\mathcal{A}} \ J_{\mathcal{A}  \mathcal{B}}\,.
\eea
The first term gives the sharp twist in $A$ while the second term gives a  smearing term in the buffer zone $Z=(A\cup B)'$.\footnote{We will discuss in section \ref{charges} non-trivial contributions that may come from the boundary $x\in \partial A$. } 
 
To understand the action of the twist suppose $U(a)$ is a translation by a vector $a$ in space-time. We want to follow the action of the translation twists $\tau(a)$ on field operators $\psi(x)$, $x\in A$, for increasing values of the translation parameter $a$. Eqs. (\ref{rr},\ref{rr1}) allow us to understand the action of the twist on the field whenever $x\in A$ and $x+a\in A$
\be
\tau(a)^\dagger \,\psi(x)\, \tau(a)=\psi(x+a)\,,\hspace{.7cm} x\in A\,,\,\, x+a\in A\,. \label{18}
\ee
In such case, it is just the ordinary translation operation and this holds even if $x$ and $x+a$ belong to disjoint connected components of $A$. When, on the contrary, $x+a \in A'$, we have, using (\ref{eq:deftwist}), (\ref{prop}), (\ref{qq}),
        \be
            \tau(a)^\dagger \ \psi(x) \ \tau(a) = 
            J_{\mathcal{A}  \mathcal{B}} \ J_{\mathcal{A}} \  \psi(x+a)  \ J_{\mathcal{A}} \ J_{\mathcal{A}  \mathcal{B}}\,, \hspace{.7cm} x\in A, \,\, x+a\in A' \ \text{.}
        \label{eq:WsobreelementofueraHxH}
        \ee
This indicates that if the global translation takes the field to the complementary spatial region $A'$, then the twist takes the operator to $J_{\mathcal{A}  \mathcal{B}}\,\mathcal{A}  \, J_{\mathcal{A}  \mathcal{B}}\subset {\cal N}$. This never leaves ${\cal N}$ but is localized in the algebra corresponding to the ``buffer zone'' $Z=(A\cup B)'$ outside both $A$ and $B$.\footnote{For a more general case where $x+a$ is not in the causal region corresponding to $A$ nor in $A'$, we have first to decompose $\psi(x+a)$ into fields in $A$ and $A'$ using the equations of motion. We will not need this decomposition in this paper because we focus on a chiral field on the line.}

These equations show that to compute a standard twist explicitly we need to understand the action of the modular conjugations $ J_{\mathcal{A}}, J_{\mathcal{B}},  J_{\mathcal{A}  \mathcal{B}}$. In general, it is difficult to understand these operators explicitly, especially in the case of  $J_{\mathcal{A}  \mathcal{B}}$ which corresponds to regions with more than one connected component. Up to the present day, these types of operators are known explicitly only for the chiral fermion. We now turn to this case.

\section{Modular conjugation for the free chiral fermion}
\label{chiral}

For fermions there is a slight modification to be made to the modular operators. If ${\cal A}$ is the algebra of a region containing the fermion field $\psi(x)$, ${\cal A}'$ will not contain the fermion field in the complementary region because fermions anticommute. This can be corrected in a standard way by a unitary transformation \cite{wassermann1998operator}. 
Define the operator fermionic sign $\Gamma$ by 
\begin{equation}
    \Gamma^{2} = 1 \text{,} \quad
    \Gamma^{\dagger} = \Gamma \text{,} \quad
    \Gamma \psi(x) \Gamma = - \psi(x) \ \text{,}
    \label{eq:def_gamma}
\end{equation}
and from it the unitary $Z$ 
\begin{equation}
    Z = \frac{1 - i \Gamma}{1 - i} \text{.} 
\end{equation}
We have
\begin{equation}
    Z Z^{\dagger} = 1 \text{,} \quad
    Z \psi(x) Z^{\dagger} = -i \Gamma \psi(x) \text{,} \quad
    Z \psi(x) \psi(y) Z^{\dagger} = \psi(x) \psi(y)  \ \text{.}
    \label{eq:def_Z}
\end{equation}
For an even fermion state, $\Gamma \ket{\Omega} = Z \ket{\Omega} = \ket{\Omega}$, we also have
\begin{equation}
    \Gamma J \Gamma = J  \text{,} \quad
    JZ = Z^{\dagger} J \ \text{.}
\end{equation}
Defining the twisted modular reflection
\begin{equation}
    \widetilde{J} = ZJ \ \text{,}
\end{equation}
this operator maps the fermion algebra ${\cal A}$ into the twisted commutant $\tilde{{\cal A}}$ containing all even operators that commute with ${\cal A}$ and all odd operators that commute with the even part of ${\cal A}$ and anticommute with the odd part of ${\cal A}$. As a result, understanding commutation as graded commutativity, and commutants as twisted ones, the twisted modular reflection plays the role of the ordinary Tomita-Takesaki modular reflection in fermionic theories. However, in the construction of twists of the preceding section, this modification plays no role since the symmetries are represented by even operators with even charge densities, where the actions of $J$ and $\widetilde{J}$ coincide. In other terms, in formulas like  (\ref{eq:WsobreelementofueraHxH}), where two modular reflections are applied to a fermion operator, the use of $J$ or $\widetilde{J}$ is indifferent. Nevertheless, to study the action of the modular reflection on a fermion field it will be convenient to use $\widetilde{J}$. 

We will focus on the chiral Majorana fermion model. This is a field $\psi(x)=\psi(x)^\dagger$, $x\in \mathbb{R}$, with anticommutation relation 
\be
\{\psi(x),\psi(y)\}= \delta(x-y)\,,
\ee   
and Hamiltonian 
\be
H=i/2 \int dx\, :\psi(x)\partial_x \psi(x):\,.\label{tmunu}
\ee
The vacuum is Gaussian with two-point function 
\be
\langle \Omega| \psi(x)\, \psi(y)|\Omega\rangle=(2 \pi i)^{-1}(x-y-i \epsilon)^{-1}\,.
\label{eq:correlador}
\ee
The coordinate $x$ can be thought of as a null coordinate in two dimensions. The Hamiltonian is positive definite and translates fields in $x$.

The modular operators for the chiral fermion are known. For a Gaussian state, the modular Hamiltonian is quadratic on the fields. The corresponding kernel was diagonalized in \cite{casini2009reduced}, obtaining the modular flow. This was analyzed in detail  \cite{rehren2013multilocal,arias2018entropy,longo2010geometric,
hollands2021modular,bueno2020reflected,erdmenger2020resolving,
wong2019gluing,tedesco2014modular},
and generalized in several works \cite{blanco2019modular,fries2019entanglement,Chen:2022nwf}.  
The modular conjugation was computed in \cite{abate2023modular} (see also \cite{longo2018relative,bueno2020reflected,mintchev2022modular}).
We  define a conjugated fermion field as
\begin{equation}
    \widetilde{\psi}(x) =  \widetilde{J} \psi(x) \widetilde{J} \ \text{.}
\end{equation}
Since the theory is Gaussian the action of $\widetilde{J}$ transforms linearly the field to the adjoint field (the modular flow also acts linearly for these Gaussian fields). For the convenience of the reader, we re-derive this linear action of $\tilde{J}$ from the knowledge of modular flow in the appendix \ref{mcmf}, in a slightly different setup as in \cite{abate2023modular}. 
 
To express the result consider a region $A$ formed by $n$ disjoint intervals $A=(a_1,b_1) \cup (a_2,b_2)\cup \cdots \cup (a_n,b_n)$, where the intervals are ordered from left to right. The complementary region $A'=(b_1,a_2)\cup (b_2,a_3)\cup\cdots \cup (b_n,\infty)\cup (-\infty,a_1)$, can also be thought to be formed by $n$ intervals in the compactified line, where $(b_n,a_1)\equiv (b_n,\infty)\cup (-\infty,a_1)$. 
Define the functions
\begin{equation}
   \Pi_{a}(x) = \Pi_{i=1}^{N} (x-a_{i})  \,, \quad
    \Pi_{b}(x) = \Pi_{i=1}^{N} (x-b_{i}) \,, \quad f(x)=-\frac{\Pi_a(x)}{\Pi_b(x)}\,. \label{productos}
\end{equation}
$f(x)$ is always increasing. It is positive in $A$ and increases from $0$ to $\infty$ in each interval. On the other hand, in the complement $A'$, $f(x)$ is negative and increases from $-\infty$ to $0$ in each interval.  Then, the equation
\be
f(x)=-f(y)\,,\label{equi}
\ee
gives, for each $y\in A$, $n$ different real solutions for $x$ 
\be
x_j=s_j(y)\,,\qquad j=1,\cdots, n\,.
\ee
Each of the solutions $s_j(y)$ belongs to a different interval in the complement $A'$ of $A$. The functions $s_j(y)$ are decreasing with $y\in A$, i.e.,  $s_j'(y)<0$. Note the same functions $s_j$ solve the equation (\ref{equi}) for the complementary region $A'$. 

Using these functions, the modular conjugation  $\widetilde{\psi}(y) =  \widetilde{J} \psi(y) \widetilde{J}$ (we have $\widetilde{J}^\dagger=\widetilde{J}$) is expressed as \cite{abate2023modular} 
\begin{equation}
    \widetilde{\psi}(y) = \sum_{j = 1}^{N} a_j(y) \ \psi(s_j(y)) \ \text{,}
    \label{eq:JpsiJ_formula1}
\end{equation}
where 
\be
  a_j(y) =  \frac{2 \,f(s_j(y)) }{f'(s_j(y))\,(y-s_j(y))}=\frac{2}{(y-s_j(y))}\left(\sum_{i=1}^n \frac{1}{s_j(y)-a_i}-\frac{1}{s_j(y)-b_i}\right)^{-1} \,.  \label{mate}
\ee
This is valid for any $y$, not necessarily in $A$. The modular conjugation is the same for the algebras $A$ and $A'$. This is reflected here in that the functions $s_j(y)$ are the same for $A$ and $A'$ and $f$ is transformed into $ -1/f$ for $A'$. This leaves (\ref{mate}) unchanged. 

Equation (\ref{eq:JpsiJ_formula1}) has to be understood as an equation for distributions in the variable $y$. It gives the modular transformed field operator as a discrete sum of field operators. Each of the $n$ operators in this sum is localized in one interval of the complementary region $A'$. How we can understand the relative weight of each of these $n$ components? To answer this question recall that as 
 $\widetilde{J}$ is an antiunitary operator 
\begin{equation}
    \{ \widetilde{\psi}(x), \widetilde{\psi}(y) \}
    = \widetilde{J} \{ \psi(x), \psi(y) \} \widetilde{J} 
     =\delta(x - y) \ \text{.}
    \label{eq:anticonmutador_1}
\end{equation}
On the other hand, we have from (\ref{eq:JpsiJ_formula1})
\begin{equation}
\begin{split}
    &\{\widetilde{\psi}(x), \widetilde{\psi}(y)\} = 
    \sum_{j, i = 1}^{N} a_i(x) a_j( y) \ \delta(s_i(x) - s_j(y)) \\
    &= \left(\sum_{i = 1}^{N}  \frac{a_i(x)^2}{|s'_i(x)|}\right) \delta(x - y) +
    \sum_{j, i = 1, \ i \neq j}^{N} a_i(x) a_j(y) \ \delta(s_i(x) - s_j(y))
    \ \text{.}
    \label{eq:anticonmutador_2}
\end{split}
\end{equation}
 Comparing (\ref{eq:anticonmutador_1}) and (\ref{eq:anticonmutador_2})
we arrive at
\be
   \sum_{i = 1}^{N}  \frac{a_i(x)^2}{| s'_i(x)|}= 1 \ \text{,}
    \label{eq:condicion_C_1}
 \hspace{1cm}  \sum_{j, i = 1, \ i \neq j}^{N} a_i(x) a_j(y) \ \delta(s_i(x) - s_j(y)) = 0 \ \text{.}
\ee
From (\ref{eq:JpsiJ_formula1}) and (\ref{eq:condicion_C_1}) we interpret 
\begin{equation}
    P_{i}(x) = \frac{a_i(x)^2}{| s'_i(x)|} 
    \label{eq:peso_psi}
\end{equation}
as the proportion of the field at the point $s_{i}(x)$ in the decomposition of $\widetilde{\psi}(x)$ in local operators. In fact, $a_i(x)/|s_i'(x)|^{1/2}$ can be thought of as the true amplitude of the different components when the fields are normalized to have a unit anticommutator.  The second relation in (\ref{eq:condicion_C_1}) implies algebraic relations satisfied by the coefficients of the different components of the conjugated field.

The formulas (\ref{eq:JpsiJ_formula1}) and (\ref{mate}) give the modular conjugation. To make these formulas more explicit we only need the roots $s_j(y)$ of eq. (\ref{equi}). These are polynomial equations of order $n$. For one interval $A=(a,b)$ we have a single root 
\begin{equation}
    s(y) = \frac{ab - \frac{(a + b)}{2} y}{\frac{(a + b)}{2} - y}  \,.
    \label{eq:sol_x_1_intervalo}
\end{equation}
This is a conformal reflection mapping the interior of the interval into the exterior. The conjugation of the field gives 
\begin{equation}
    \widetilde{\psi}(y) = \frac{b - a}{a + b - 2y} \psi(s(y)) \ \text{.}
    \label{eq:unintervalo_sol_1}
\end{equation}
For two symmetric intervals $A = (-b, -a) \cup (a, b)$ the two solutions, that we can call $s_{\pm}$, are
\be
s_{\pm}(y) = \frac{-(a - b)^2 y \pm \sqrt{(a - b)^4 y^2 + 4  a b (y^2 - ab)^2}}{2 (ab - y^2)}\,.
\label{eq:sol_x_2_intervalo}
\ee

\subsection{Fermions on the circle}
\label{fc}
It is also useful to consider the compact picture of the line by conformally mapping the line to the circle of the unit modulus complex numbers $|z|=1$. This picture allows us to obtain simpler expressions for some specific symmetric regions. See \cite{rehren2013multilocal,
hollands2021modular}. 
The mapping is given by Cayley transform
\begin{equation}
   z = \mathrm{Ca}(x) = \frac{1 + i x}{1 - i x} \ \text{.}
\end{equation}
Under the Cayley transformation, the real fermion field transforms as
\begin{equation}
    \check{\psi}(z) = \frac{1}{\sqrt{ -i \mathrm{Ca}^{\prime}(x)}} \psi(x) = \frac{i\,(i+x)}{\sqrt{2}} \psi(x) \text{,}
\label{eq:def_psi_complejo} 
\end{equation}
\noindent where $z = \mathrm{Ca}(x)$. Then one has the conjugate field
\begin{equation}
    \check{\psi}^{\dagger}(z)
    = z\, \check{\psi}(z) \,.
\end{equation}
These fields have the anti-commutation rule
\begin{equation}
    \{\check{\psi}(z), \check{\psi}^{\dagger}(w) \} = \frac{1}{| \mathrm{Ca}^{\prime}(x) |} \delta(x - y) := \check{\delta}(z - w)
    \, . 
\label{eq:conmutador_phi_check_1}
\end{equation}

To construct symmetric regions let $I = (U, V)$ be an arc of $S^{1}$. Call
$\sqrt[n]{I} = \bigcup_{j=1}^{n} (u_{j}, v_{j})$ the set of $n$
symmetrical arcs distributed on $S^{1}$ that arises from applying the n-root to the points of $I$.
If we applied to these arcs the inverse Cayley transform we obtain a distribution
of intervals on the real axis $A = \mathrm{Ca}^{-1}(\sqrt[n]{I}) = \bigcup_{j=1}^{n} (a_{j}, b_{j})$.

Now if we want to study the modular conjugation in this distribution of intervals $A$
we have to use the function defined on (\ref{productos}) and the condition (\ref{equi}).
It can be shown that 
\begin{equation}
    f(x) = \prod_{j=1}^{n} \frac{x-a_{j}}{x- b_{j}} =
    \text{cons.} \ \prod_{j=1}^{n} \frac{z - u_{j}}{z - v_{j}} =
    \text{cons.} \ \frac{z^{n}-U}{z^{n}-V} =
    \check{f}(z) \, ,
    \label{eq:longo_prop_5}
\end{equation}
 where we used $z = \frac{1 + i x}{1 - i x}$, $u_{j} = \frac{1 + i a_{j}}{1 - i a_{j}}$,
$v_{j} = \frac{1 + i b_{j}}{1 - i b_{j}}$, $U= u_{j}^{n}$, $V= v_{j}^{n}$.
The condition $f(x) + f(y) = 0$ is equivalent to 
$\check{f}(z) + \check{f}(w) = 0$, with $z, w \in S^{1}$. This writes
\begin{equation}
    w^{n} = \frac{\frac{U + V}{2} z^{n} - U V}{z^{n} - \frac{U + V}{2}} \, .
    \label{eq:solucion_n_intervalos_complejo}          
\end{equation}
Taking the n-roots of $w^{n}$ we obtain the $n$ solutions $\check{s}_{j}(z)$ that we want.
It will be useful to rename the solution for one interval as
\begin{equation}
    s_{0}(z) =  \frac{\frac{U + V}{2} z - U V}{z - \frac{U + V}{2}}\, ,
\label{eq:s_0_definicion}
\end{equation}
The solutions $s_{j}(z)$ for $n$ intervals then obey
\begin{equation}
    s_{j}(z)^{n} = s_{0}(z^{n}) \, .
\label{eq:identidad_1_sj}
\end{equation}

Now we want to study the modular conjugation on these fields. We define
\begin{equation}
    \check{\widetilde{\psi}}(z) =  \widetilde{J} \check{\psi}^{\dagger}(z) \widetilde{J} \, .
\label{eq:def_JpsiJ_complejo}
\end{equation}
Using (\ref{eq:JpsiJ_formula1}) and (\ref{mate}) we can show that 
\begin{equation}
    \check{\widetilde{\psi}}(z) = \sum_{j = 1}^{n} d_{j}(z) \check{\psi}(\check{s}_{j}(z)) \, ,
\label{eq:JphiJ_complejo}
\end{equation}
where 
\begin{equation}
    d_{j}(z) = \frac{\check{f}(\check{s}_{j}(z))}{\check{f}^{\prime}(\check{s}_{j}(z))} \frac{2}{z - \check{s}_{j}(z)} \ \text{.}
\label{eq:dj_expresion_2}
\end{equation}
We can see that $d_{j}(z)$ conserves the same expression that had $a_{j}(x)$.
Using  (\ref{eq:longo_prop_5}) and (\ref{eq:identidad_1_sj})
we find, more explicitly,
\begin{equation}
    d_{j}(z) = \frac{2}{n (U - V)} \, \frac{(s_{0}(z^{n}) - U)(s_{0}(z^{n}) - V)}{s_{0}(z^{n})} \, \frac{s_{j}(z)}{ z - s_{j}(z)} \, ,
\label{eq:dj_expresion_explicita_simetrico}
\end{equation}
where $s_{0}(z)$ was defined in (\ref{eq:s_0_definicion}).

To compute the weights $\check{P}_{j}(z)$ of each component we follow
the same procedure as above. First, we check the anti-commutation
relation of $\check{\widetilde{\psi}}(z)$,
\bea
    &&\hspace{-1cm} \{ \check{\widetilde{\psi}}(z) , \check{\widetilde{\psi}^{\dagger}}(w)\}
    =  \widetilde{J} \{ \check{\psi}^{\dagger}(z) , \check{\psi}(w)\} \widetilde{J}
    = \check{\delta}(z - w) \nonumber \\    
    &&\hspace{2cm}= \sum_{i = 1}^{n} \frac{|d_{i}(z)|^{2}}{|\check{s}_{i}^{\prime}(z)|} \check{\delta}(z - w) + 
        \sum_{j, i = 1, \ i \neq j}^{n} d_{i}(z) \left(d_{j}(w)\right)^{*} \check{\delta}(\check{s}_{i}(z) - \check{s}_{j}(w))
    \,.
\label{eq:conmudator_phi_check_widel}
\eea
The second term of this last expression vanishes. Then the weights are  
\begin{equation}
    \check{P}_{j}(z) = \frac{|d_{j}(z)|^{2}}{|\check{s}_{j}^{\prime}(z)|} = \frac{|a_{j}(x)|^{2}}{|s_{j}^{\prime}(x)|} = P_{j}(x) \,.
\end{equation}
It is interesting to emphasize that the weights $P_{i}(x)$ do not transform 
under the Cayley transformation, as expected. Using the identity 
\begin{equation}
    \check{s}_{j}(z)^{n} = s_{0}(z^{n})
    \Rightarrow
    \check{s}^{\prime}_{j}(z) = \frac{s_{0}^{\prime}(z^{n})}{\check{s}_{j}(z)^{n - 1}} z^{n - 1} 
    \ \text{,}
\end{equation}
we can calculate the weights $\check{P}_{j}(z)$ as
\begin{equation}
    \check{P}_{j}(z) = \frac{|d_{j}(z)|^{2}}{|\check{s}_{j}^{\prime}(z)|}
    = \frac{4}{n^{2}}\, \left| \frac{(s_{0}(z^{n}) - U)^{2} (s_{0}(z^{n}) - V)^{2}}{s^{\prime}_{0}(z^{n}) \ (U - V)^2}\right| \, \left|\frac{1}{(z - \check{s}_{j}(z))^{2}}  \right| \ \text{.}
\label{eq:Pj_expresion_explicita_simetrico}
\end{equation}
We can see that for the chosen special configuration of regions, both, $d_{j}(z)$ and $\check{P}_{j}(z)$, have a factor that does not
depend on the solution $\check{s}_{j}(z)$, times a simple term that depends on the particular solution. 

\section{Standard translation twists}
\label{explicit}

In this section we give examples of standard twists for one and two intervals, computing the action of the one-parameter groups of translations on the field operators. For one interval, we compare with twists formed by smearing the Noether charge. For two intervals, we display how the standard twists make the fields jump from one interval to the other, without disturbing the outside region. 

\subsection{Standard twist vs Noether twist for one interval}
\label{S:standar_noether_twist}

\begin{figure}[H] 
     \centering 
     \makebox[\textwidth][c]{\includegraphics[width=0.6\textwidth]{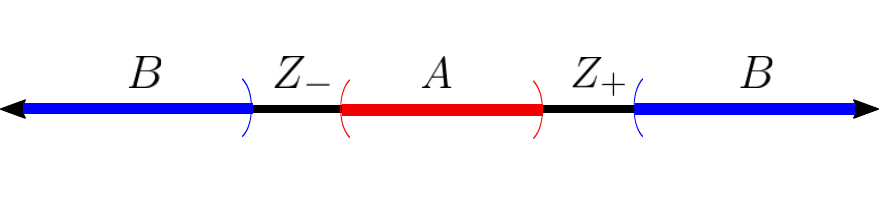}} 
     \caption{Distribution of intervals on the line.} 
     \label{fig:twist_dos_intervalos_centrado_sinnumeros} 
\end{figure}

Let us take the interval $A=(-a,a)$ and the smearings zones $Z_{-} = (-\epsilon-a,-a)$ and $Z_{+} = (a,a+\epsilon)$ which together form $Z=Z_{-} \cup  Z_{+}$.
Hence, the complementary region is $B=(-\infty,-a-\epsilon)\cup (a+\epsilon,\infty)$.

Let us first construct a twist by smearing the Noether charge. The local charge  has the generic form
\begin{equation}
  H_\alpha= \frac{i}{2} \int dx \ \alpha(x)\, :\psi(x)\,\partial_x \psi(x):  \,.
  \label{eq:generador_twist_evolucion}
\end{equation}
where $\alpha(x)$ is a smooth function such that $\alpha(x)=1$ for $x\in A$ and $\alpha(x)=0$ for $x\in B$. The twist translation group is given by
\begin{equation}
  \tau_{\alpha}(t) = e^{-iH_{\alpha}t} \ \text{.}
  \label{eq:def_twist_simple}
\end{equation}  
Writing the transformed field as 
\begin{equation}
\hat{\psi}(x, t) = \tau_\alpha(t)^\dagger \psi(x) \tau_\alpha(t) \,,
\end{equation}
we obtain, deriving with respect to the time parameter,
\begin{equation}
  \frac{d \hat{\psi}(x, t)}{d t} =
  i \left[H_\alpha, \hat{\psi}(x,t)  \right] = \alpha(x) \partial_x\hat{\psi}(x, t) + 
          \frac{1}{2}\partial_x \alpha(x) \hat{\psi}(x, t)\,.
\label{eq:EDP_1}
\end{equation}
Assuming $\alpha(x)> 0$ for $x\in A\cup Z$ for simplicity, the solution of this differential equation with the initial condition $\hat{\psi}(x,0)=\psi(x)$,  $x\in A$, is
\begin{equation}
  \hat{\psi}(x, t) = \sqrt{\frac{\alpha(g^{ - 1}(t + g(x)))}{\alpha(x)}}\,\,\psi(g^{ - 1}(t + g(x))) \,, \quad g(u) = \int_{x_{0}}^{u} \frac{1}{\alpha(s)} \ ds \,.
  \label{eq:sol_general_psi}
\end{equation}
$x_0$ is an arbitrary point inside $A$. The formula for $\hat{\psi}(x, t)$ is in fact independent of $x_0$. As $g$ and its inverse function $g^{-1}$ are increasing, the action of the twist for positive parameter $t$ always moves the field to the right. In addition, the field gets multiplied by a normalizing factor that is necessary to keep invariant anti-commutators. Hence, while initially, for small enough $t$, the twist just translates inside $A$, for larger positive $t$, it squeezes all operators to the right interval $(a, a+\epsilon)$ of the buffer zone, and towards the point $a+\epsilon$ in the limit $t\rightarrow \infty$. For negative values of $t$ the fields are moved in the opposite direction, and end up squeezed on the left buffer zone for large negative $t$.    

\begin{figure}[t] 
     \centering 
     \makebox[\textwidth][c]{\includegraphics[width=0.5\textwidth]{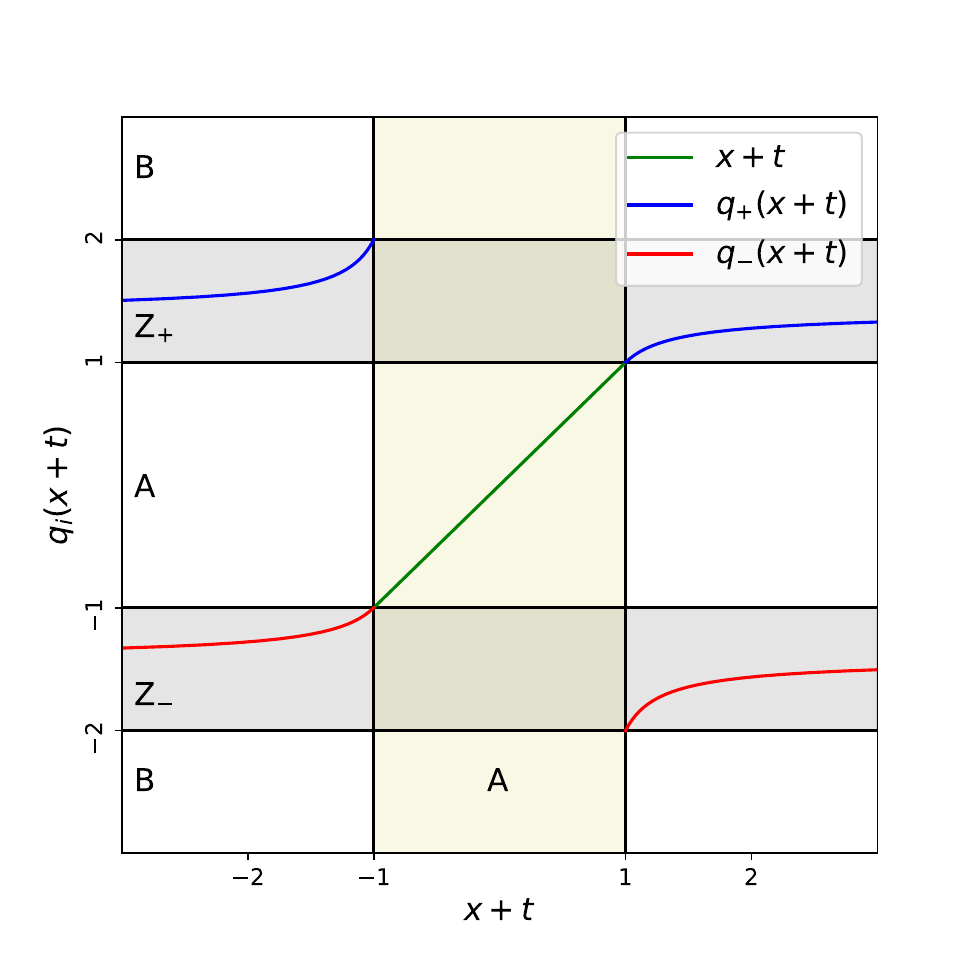}} 
     \caption{The twists moves the field as an ordinary translation if $x+t\in A$ (green dashed line). Afterward, for $x+t\in A'$, it splits it into two fields at the positions of $q_{+}(x + t)$ and $q_{-}(x + t)$ inside the buffer zone. This is shown here by the blue and red curves, where we have taken $a=1$ and $\epsilon=1$.}
     \label{fig:posiciones_2intervalos_centrado} 
\end{figure}

Now we describe the action of the standard translation twists. Let us start with a point $x\in A$ and apply a twist $\tau_A(t)$ of translations of parameter $t$. Eq. (\ref{18}) and (\ref{eq:WsobreelementofueraHxH}) give
\begin{equation}
    \tau_{A}(t)^\dagger \psi(x) \tau_{A}(t) =
    \left\{\begin{array}{lcc}
        
        \psi(x+ t)\,, &  & x + t \in A \\
        J_{\mathcal{AB}} \ J_{\mathcal{A}} \  \psi(x + t) J_{\mathcal{A}} \ J_{\mathcal{AB}}\,, &  & x+ t \notin A \\
    \end{array} \right. \,. \label{62}
\end{equation}
The composition of the two modular conjugations is computed with the formulas of section \ref{chiral}. $J_{\mathcal{A}}$ corresponds to a single interval and is associated with the mapping $x\rightarrow s(x)$, where $s(x)$ follows form (\ref{eq:sol_x_1_intervalo}) by replacing  $a\rightarrow -a, b\rightarrow a$:
\be  
s(x)= \frac{a^2}{x} \,,\hspace{.7cm} \widetilde{J}_{\cal A} \psi(x) \widetilde{J}_{\cal A}= \frac{a}{x}\, \psi(a^2/x)\,.
\ee
This point transformation is composed with the one effected by $J_{\mathcal{AB}}$ that takes a point and sends it to two possible locations according to the functions $s_\pm$ of (\ref{eq:sol_x_2_intervalo}), where we have to take $b=a+\epsilon$. As a result, the composition  $J_{\mathcal{AB}} \ J_{\mathcal{A}}$ maps a field at a point $x$ to two possible positions given by
\bea
q_\pm(x)=s_\pm(s(x))=\frac{x \left(-a^2 \epsilon^2 \pm x \sqrt{a^3 \left(4 (a+\epsilon) \left(-\frac{a^3}{x^2}+a+\epsilon\right)^2+\frac{a \epsilon^4}{x^2}\right)}\right)}{2 a x^2 (a+\epsilon)- 2 a^4}\, \label{eq:qpm}.
\eea
This is shown in figure \ref{fig:posiciones_2intervalos_centrado}. When $x+t\in A$, the twist just performs the translation by $t$. For $x+t \in A'$ the action of the twist is non-local, and the field is split into two components at the points $q_\pm(x+t)=s_\pm(s(x+t))$, each one in of the intervals $(-\epsilon-a, -a), (a,a+\epsilon)$ of the buffer zone $Z$. It never leaves the buffer zone in this case. In the limit $t\rightarrow \pm \infty$ the position of the two fields gets frozen at $q_\pm=\pm \sqrt{a(a+\epsilon)}$. 

\begin{figure}[t] 
     \centering 
     \makebox[\textwidth][c]{\includegraphics[width=0.5\textwidth]{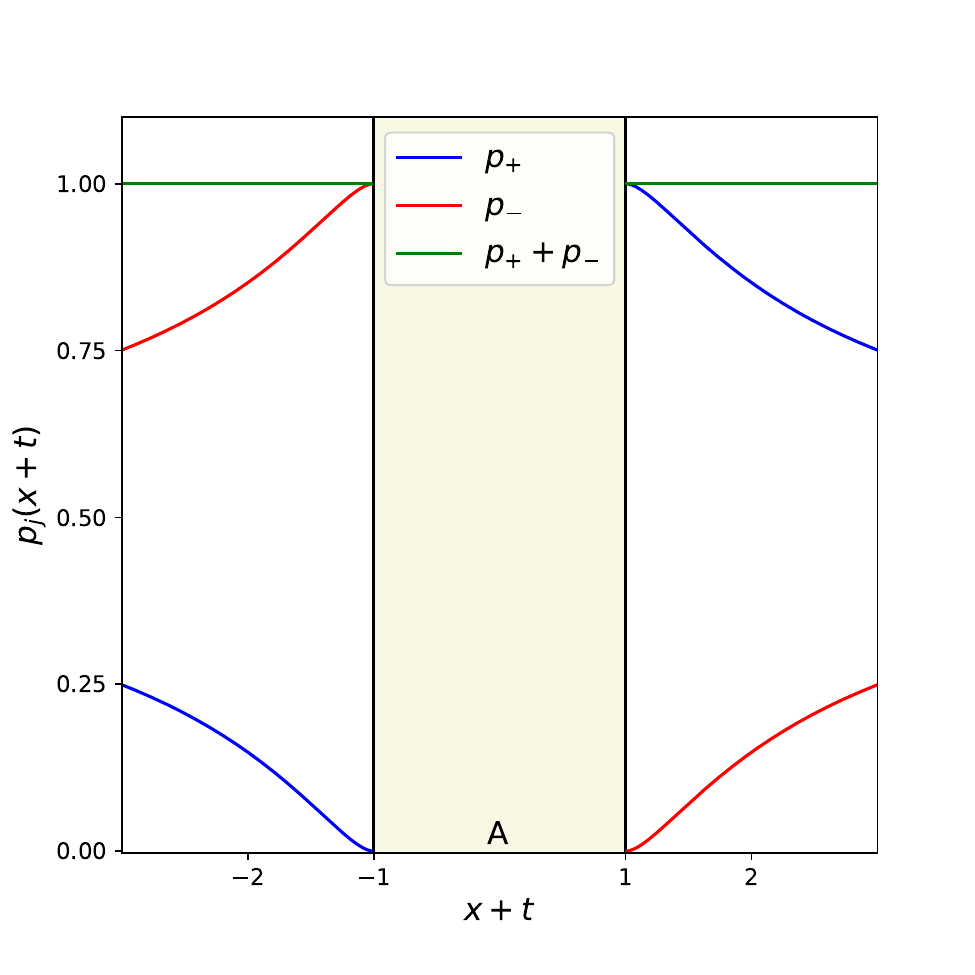}} 
     \caption{The weights of both components of the modular conjugated field are shown here by the red and blue curves. The geometry is the same as in figure \ref{fig:posiciones_2intervalos_centrado}, and the colors of the curves here correspond to the trajectories displayed there.} 
     \label{fig:probabilidades_2intervalos_centrado} 
\end{figure}

More precisely, the transformation is
\be
\tau_{A}(t)^\dagger \psi(x) \tau_{A}(t)= c_{+}(x+t) \, \psi(q_+(x+t))+ c_{-}(x+t) \,\psi(q_-(x+t))\,,  \hspace{.7cm} x+t\in A'\,,
\label{eq:twist_estandar_1}
\ee

\noindent where

\begin{equation}
    c_{\pm}(x) = \frac{a}{x} \frac{1}{q_{\pm}(x)-\frac{a^2}{x}} \left(\frac{(a + \epsilon)}{(a + \epsilon)^{2} - q_{\pm}(x)^{2}} + \frac{a}{q_{\pm}(x)^{2} - a^2} \right)^{-1} \ \text{.}
\label{eq:c_pm}
\end{equation}

\noindent The coefficients $c_{\pm}$ were obtained from the use of (\ref{mate}) for $J_{\mathcal{A}}$ and next for $J_{\mathcal{AB}}$ and finally replacing $a_ 1=-(a+\epsilon), b_ 1=-a, a_ 2=a, b_ 2=a+\epsilon$.

The probabilities associated with the two components are
\be
p_\pm(x+t)=\frac{c_{\pm}(x+t)^{2}}{|q'_\pm(x+t)|} \,,\hspace{.8cm} p_{+}(x+t)+p_{-}(x+t)=1 \,.
\ee

They are plotted in figure \ref{fig:probabilidades_2intervalos_centrado}.
When $x+t\rightarrow a$ we have $q_+(a)=a$ and $q_-(a)=-(a+\epsilon)$.
At that point, $p_+(a)=1$ and $p_-(a)=0$, giving the continuity of the operator as it crosses from $A$ to the buffer zone,
and is split in two for larger $t$. In the same way, for $x+t=-a$, $q_-(-a)=-a$, $q_+(-a)=a+\epsilon$, and $p_+(-a)=0$, $p_-(-a)=1$. 

For large $|x+t|\rightarrow \infty$, the limit positions $q_{\pm}(\pm \infty)=\pm \sqrt{a(a+\epsilon)}$ on the two buffer intervals get equal probabilities $p_\pm(\pm \infty)=1/2$ is this symmetric case. This is in contrast with the Noether twist where the fields get squeezed to $\pm (a+\epsilon)$ for $x+t\rightarrow \pm \infty$. However, if we choose the two buffer zones of different sizes, say $\epsilon_1$ and $\epsilon_2$ for the left and right intervals, and take a limit of $\epsilon_1\ll \epsilon_2, a$, we get a vanishing probability for the left component in the large $t$ limit:
\be
\lim_{t\rightarrow \infty} p_-(x+t) \sim \frac{8 (a + \epsilon_{2})^{2} (2 a + \epsilon_{2})}{a \epsilon_{2} (4 a + 3 \epsilon_{2})^{2}} \epsilon_{1} + O(\epsilon_{1}^{2}) \,.
\ee

It is interesting to remark that the action of the twist is continuous but not infinitely differentiable, in the sense, for example, of its action on the correlation function
\be
\langle\Omega | \tau_{A}(t)^\dagger \psi(x) \tau_{A}(t)\,\psi(y)|\Omega\rangle\,.
\ee
This function has discontinuous first derivatives as a function of $t$ or $x$ when $x+t$ hits the buffer zone. This is analogous to the discontinuity of the second derivative in the density of the type I factor ${\cal N}$ found in \cite{bueno2020reflected}. In contrast, we can make the action of the Noether twist  (\ref{eq:sol_general_psi}) infinitely differentiable just by choosing $\alpha$ to be so. 

We will study the generator of the one interval standard twist in section \ref{charges}. This generator is positive definite, in contrast to the generator  (\ref{eq:generador_twist_evolucion}) of the Noether twist. 

\subsection{Jumping twists}

If the region $A$ has $n$ intervals, with $n>1$, the translation twist has an interesting effect. Because eq. (\ref{62}) is always valid for any $A,B$, if we start with the field $\psi(x)$ for $x\in A_1$ in one of the intervals $A_1$ of $A$, and apply a twist of parameter $t_0$ such that $x+t_0\in A_2$, for $A_2$  another interval of $A$, the twist will make the field ``jump'' between different intervals.    This cannot be done by twists constructed smearing the Noether charge.  It is interesting to understand how this jump is realized if we continuously change the twist parameter from $t=0$ to $t=t_0$. As the twist belongs to the algebra of $A\cup Z$, the transformed field can never have support on the region $B$ complementary to $A\cup Z$. What actually happens is that the field is moved inside $A_1$ until it reaches the buffer zone $Z$. Then it is split into different fields living in the components of the buffer zone $Z$, which consists of $2 n$ intervals separating the $n$ intervals of $A$ and the $n$ intervals of $B$. If we continue the twist translation, at some point all the components of the split field will tend to have zero weight, except one that will tend to have weight $1$. When this happens, this component enters the interval $A_2$ as a single field. The field is thus reconstituted from the buffer zone, having never passed through the intervals of $B$ that separate the different intervals of $A$. 

All this readily follows from the expressions for the twist action of the previous sections. However, in general these expressions involve the roots of a polynomial in $2 n$ variables, and are either cumbersome for $n=2$, or have to be dealt with numerically for $n>2$. For this reason, we are presenting here the case of symmetric intervals on the circle, as described in section \ref{fc}. This is the same effect, though as we are using ordinary translations in the circle, this corresponds to a particular one-parameter group of conformal transformation on the line and not to translations on the line. 

\begin{figure}[H] 
     \centering 
     \makebox[\textwidth][c]{\includegraphics[width=0.4\textwidth]{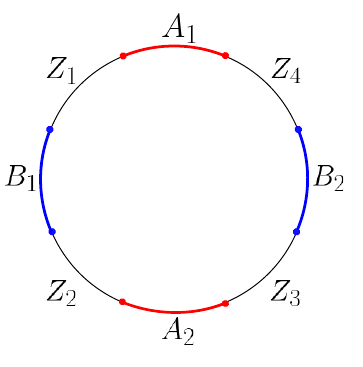}} 
     \caption{Distribution of intervals on the circle.} 
     \label{fig:4_intervalos_circulo} 
\end{figure}

\begin{figure}[t] 
    \centering 
    \makebox[\textwidth][c]{\includegraphics[width=0.6\textwidth]{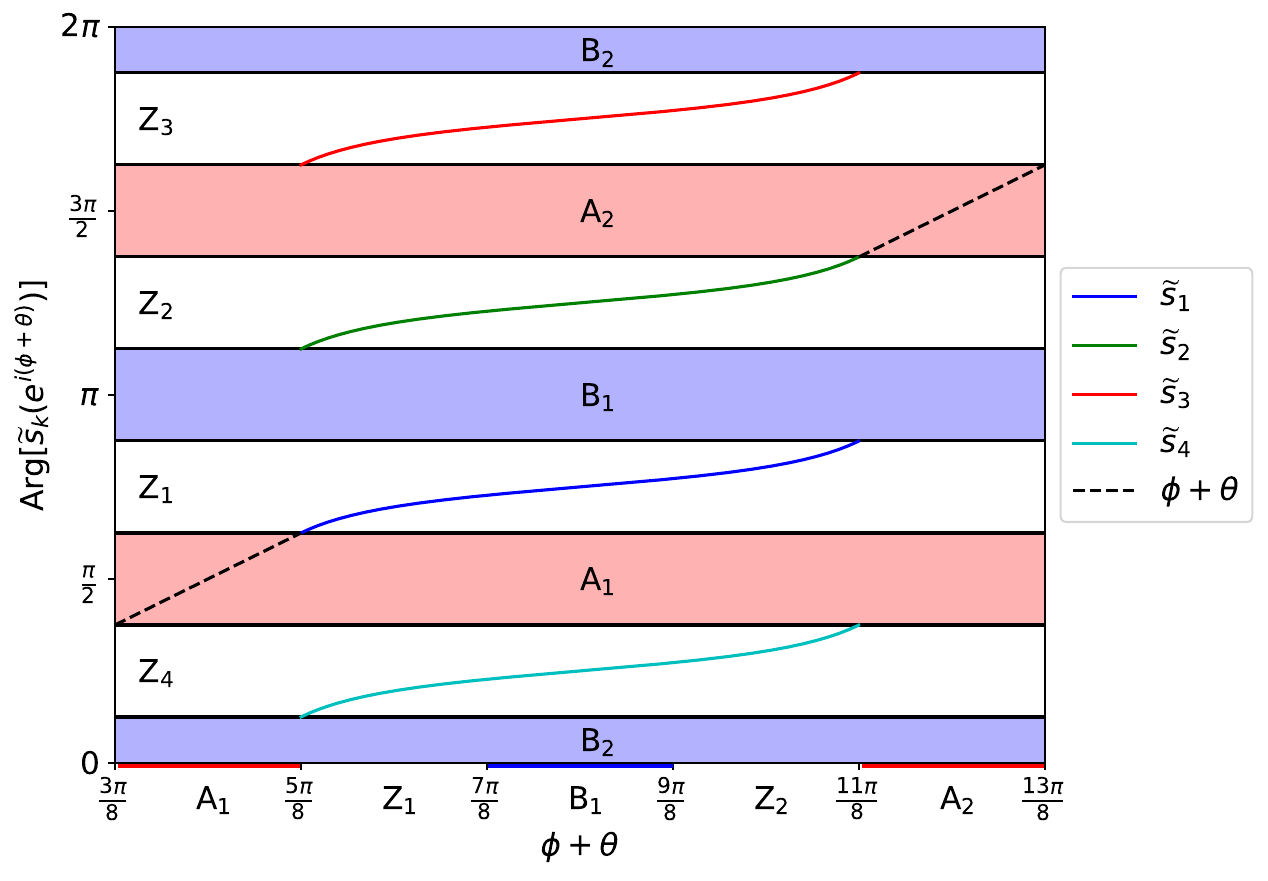}} 
    \caption{Plot of the argument of $\widetilde{s}_{k}(e^{i(\phi+\theta)})$ as a function of $\phi + \theta$
    in the intervals $A_{1} \cup Z_{1} \cup B_{1} \cup Z_{2} \cup A_{2}$, which are marked with colors in the horizontal axis.
    The horizontal strips mark the positions of the different intervals on the vertical axis.
    The black dotted line represents the case where the fermion
    is in $e^{i (\phi + \theta)} \ \in A$, where is only translated by the twist.
    The fermion is split into four branches as it leaves $A_1$.
    When it enters $A_2$, all branches disappear except one that reaches probability $1$.
     }
    \label{fig:posiciones_4_intervalos_complejo} 
\end{figure}

We take the simplest case of $n=2$ and the maximally symmetric case where we start with the interval $I_{A \cup B} = (-i, i)$ that generates $\sqrt[4]{I_{A \cup B}} = A \cup B$. This is shown in the figure \ref{fig:4_intervalos_circulo}.
Explicitly, $A= A_1\cup A_2=(e^{i \frac{3 \pi}{8}}, e^{i \frac{5 \pi}{8}}) \cup (e^{i \frac{11 \pi}{8}}, e^{i \frac{13 \pi}{8}})$
and $B= B_1\cup B_2= (e^{i \frac{7 \pi}{8}}, e^{i \frac{9 \pi}{8}}) \cup (e^{i \frac{-\pi}{8}}, e^{i \frac{\pi}{8}})$.
Also, we have $I_{A} = (e^{i \frac{3 \pi}{4}}, e^{i \frac{5 \pi}{4}})$ that generates $A = \sqrt{I_{A}}$.
Considered a fermion placed in $e^{i \phi} \in A$, then the twist that we will apply corresponds to a rotation of angle $\theta$. We have
\begin{equation}
    \tau_{A}(\theta)^\dagger \check{\psi}(e^{i \phi}) \tau_{A}(\theta) =
    \left\{\begin{array}{lcc}
        
        \check{\psi}(e^{i (\phi + \theta)})\,, &  & e^{i (\phi + \theta)} \in A \\
        J_{\mathcal{AB}} \ J_{\mathcal{A}} \  \check{\psi}(e^{i (\phi + \theta)}) J_{\mathcal{A}} \ J_{\mathcal{AB}}\,, &  & e^{i (\phi + \theta)} \notin A \\
    \end{array} \right. \,.
\end{equation}
Using the formulas developed on section \ref{fc}, we arrive at 
\begin{equation}
    J_{\mathcal{AB}} \ J_{\mathcal{A}} \  \check{\psi}(z) J_{\mathcal{A}} \ J_{\mathcal{AB}} =
    \sum_{k= 1}^{4} \left[ d_{k}(\check{g}^{A}_{1}(z)) c_{1}(z)^{*} + d_{k}(\check{g}^{A}_{2}(z)) c_{2}(z)^{*} \right] \ \check{\psi}(\widetilde{s}_{k}(z))\,,
\end{equation}
where
\begin{eqnarray}
    \check{g}^{A}_{j}(z) &=&  i (-1)^{j-1} \sqrt[2]{\frac{z^{2} + \sqrt{2}}{\sqrt{2} z^{2} + 1}} \,,
  \\  \widetilde{s}_{k}(z) &=& \check{s}^{AB}_{k}(\check{g}^{A}_{j}(z)) = e^{i\frac{\pi}{4}(2k+1)} \, \sqrt{\frac{\sqrt{2}z^{2} + 1}{z^{2} + \sqrt{2}}} \ \text{.}\label{yyy}
\end{eqnarray}
where $j=1, 2$, $k=1, 2, 3, 4$ and the principal branch is selected by the roots.  The $\check{g}^{A}_{j}(z)$ and $c_{j}(z)$ are the solutions
for the positions and amplitudes of the fermions that appear when we apply $J_{\mathcal{A}}$ and
$\check{s}^{AB}_{k}(z)$ and $d_{k}(z)$ the ones that appear when we apply $J_{\mathcal{AB}}$.
They are constructed using the formulas (\ref{eq:identidad_1_sj}) and (\ref{eq:dj_expresion_2})
using the respective $U$ and $V$ for $A$ and $A \cup B$.
We remark that $\widetilde{s}_{k}(z)$ does not depend on $j$,
this is because when we evaluate $\check{s}^{AB}_{k}(\check{g}^{A}_{j}(z))$ we have to remind the
form of the solution (\ref{eq:identidad_1_sj}). Therefore, when we perform $\check{g}^{A}_{j}(z)^{4}$
the dependence on $j$ is eliminated.
In figure \ref{fig:posiciones_4_intervalos_complejo} we can see the paths
that the four branches follow in the case where $e^{i (\phi + \theta)} \ \notin A$. 

\begin{figure}[t] 
   \centering 
   \makebox[\textwidth][c]{\includegraphics[width=0.6\textwidth]{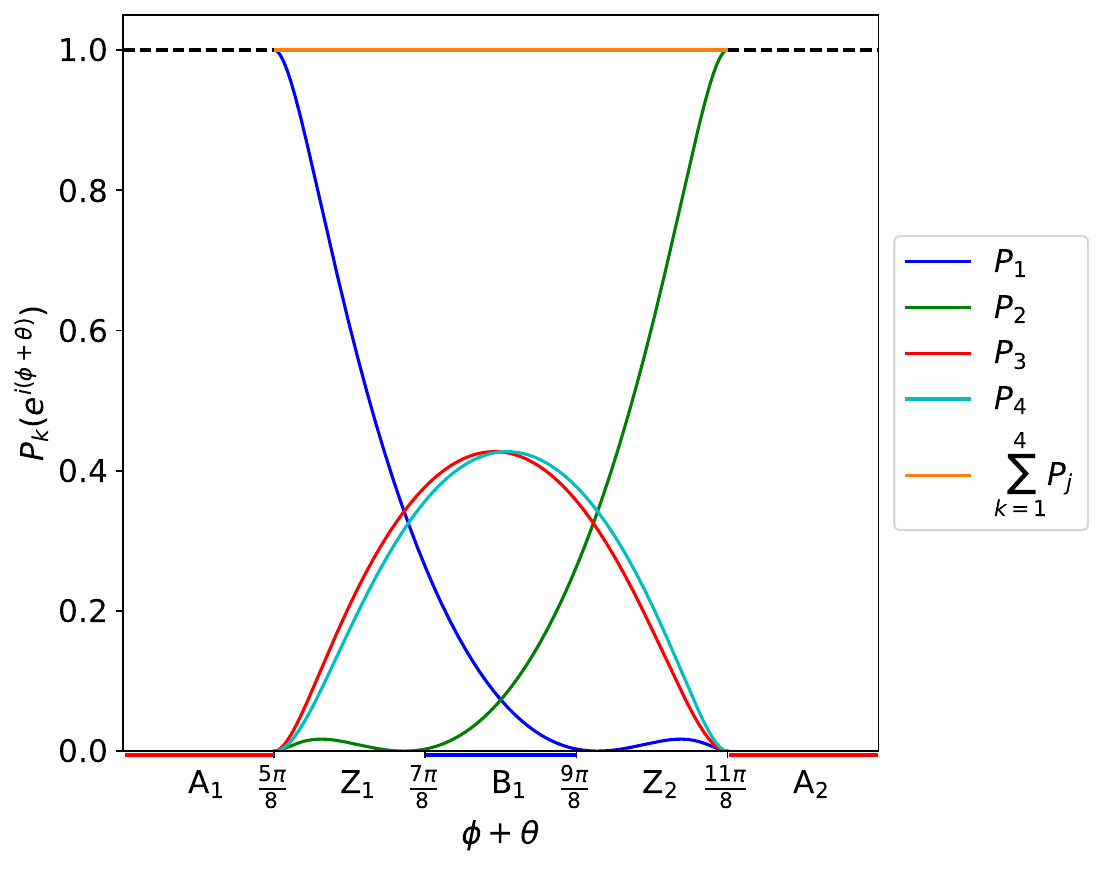}} 
   \caption{
    Plot of the probabilities $P_{k}(e^{i(\phi + \theta)})$ as a function of 
    $\phi + \theta$ in the interval $A_{1} \cup Z_{1} \cup B_{1} \cup Z_{2} \cup A_{2}$, which are marked with colors in the horizontal axis.
    The probability $P_{k}(z)$ is the one corresponding to the fermion located at $\widetilde{s}_{k}(z)$.
    The black dotted line represents the case where the fermion is in $e^{i(\phi + \theta)} \in A$, where it is not split.
    } 
   \label{fig:prob_4_intervalos_complejo} 
\end{figure}

We can calculate the probabilities following the same procedure as in
(\ref{eq:conmudator_phi_check_widel}) and obtain
\begin{equation}
   P_{k}(z) = \frac{\left| d^{AB}_{k}(\check{g}^{A}_{1}(z)) c^{A}_{1}(z)^{*} + d^{AB}_{k}(\check{g}^{A}_{2}(z)) c^{A}_{2}(z)^{*} \right|^{2}}{\left| \widetilde{s}^{\prime}_{k}(z) \right|} \ \text{.}
\label{eq:prob_4_intervalos_complejo}
\end{equation}
Is interesting to note that the probabilities are not the product 
of the probabilities that appear by acting with $J_{\mathcal{A}}$ and $J_{\mathcal{AB}}$
separately. It is because of the high symmetry of this problem that removes the dependencies of $j$ at the fermions positions (\ref{yyy}). In figure \ref{fig:prob_4_intervalos_complejo} we plot the
probabilities in the interval $A_{1} \cup Z_{1} \cup B_{1} \cup Z_{2} \cup A_{2}$, corresponding to the same range in figure \ref{fig:posiciones_4_intervalos_complejo}.

From these graphs, we can observe the following behavior. Let's assume that initially our fermion
is located in $A_{1} = (e^{i \frac{3 \pi}{8}}, e^{i \frac{5\pi}{8}}) \subset A$.
If we apply the twist $\tau^{A}(\theta)$ to this fermion, where the parameter $\theta$ is small enough to keep
it within $A$, then the twist only translates it and its movement will correspond to the black dashed line marked on the graph.
However, if the parameter is large enough to take it out of region $A$, the twist will transform the fermion into 4 fermions
located at positions $\widetilde{s}_{k}(e^{i (\phi + \theta)})$ in the four intervals $Z_{k}$ of the buffer zone.
As the parameter $\theta$ increases, the positions of the fermions will vary as seen in the graph,
until the parameter becomes large enough for $e^{i (\phi + \theta)} \in A_{2} = (e^{i \frac{11 \pi}{8}}, e^{i \frac{13 \pi}{8}}) \subset A$,
and thus we return to the solution represented by the black dashed line in the graph.
Analyzing how the probabilities behave along this trajectory, using figure \ref{fig:prob_4_intervalos_complejo},
we can see that as the fermion starts to leave the interval $A_1$, the largest probability
is $P_{1}$, which correspond to the fermion located at $\widetilde{s}_{1}(z) \in Z_{1}$.
Then the probabilities of the other fermions increase, indicating a mixture among them.
As we increase the parameter $\theta$ and $e^{i (\phi + \theta)}$ approach to $A_{2}$,
we observe that the probability $P_{1}$ starts to decrease while the probability $P_{2}$, the probability corresponding
to the fermion located at $\widetilde{s}_{2}(z) \in Z_{2}$, increases
reaching $1$ upon entering $A_{2}$.

To summarize, when the twist takes the fermion out of region $A$,
it produces a mixture of operators in the buffer zone, each with its corresponding probability $P_{k}$.
However, as we approach another interval of the region $A$ again, the mixture begins to unravel,
and we regain a single operator located in $A$. Therefore, the twist operator has successfully transported
a fermion operator from the interval $A_{1}$ to $A_{2}$ without ever being within region $B_{1}$ that separates the
intervals of $A$. In other words, it has enabled the fermion to jump from one interval to another without
ever passing through the separating spaces $B$ between the different intervals of $A$.

It is interesting to note that for translations on a circle an analogous effect of field regeneration also happens for the case where $A$ and $B$ are single intervals. In that case, the field that starts in $A$ is ordinarily translated until it hits the buffer zone on the right side of the interval, where it is converted into a non-local operator in the buffer zone.  After some more translation, the local field is regenerated and reappears on the left side of the interval $A$.

\section{Twist generator and operator bounded energy inequalities}
\label{charges}

In the previous sections we have computed how the standard translation twist acts on field operators. Now we are going to give a closed expression for the twist itself. Since twists form a continuous group, a compact way to write these twists is to display its generator $\tilde{H}$. This is defined by 
\be
\tau(x)= e^{-i \, x\,\tilde{H}}\,,  
\ee
where $x$ is the translation parameter. 
From eq. (\ref{eq:deftwist}),
\begin{equation}
    \tau(x) = W^{\dagger} \left( U(x) \otimes 1 \right) W\, \text{,}
    \label{eq:deftwist1}
\end{equation}
it follows that the generator is given by
\begin{equation}
    \tilde{H} = W^{\dagger} \left( H \otimes 1 \right) W\,,
    \label{eq:deftwist2}
\end{equation}
where $H$ is the full Hamiltonian on the theory acting on the first Hilbert space copy.  For simplicity, we will consider the specific case of the twist for a single interval treated in section \ref{S:standar_noether_twist}. The region
 where the twist acts as translations is $A = (-a, a)$, the region where it does not act is $B = (-\infty, -b) \cup (b, \infty)$ whit $a < b$, and the smearing zones are  $Z_{+} = (a, b)$ y $Z_{-} = (-b, -a)$.

For the chiral fermion the Hamiltonian writes
\begin{equation}
    H = \int_{\mathbb{R}} dx \ :T(x): 
\end{equation}
with 
\be 
T(x)= \frac{i}{2}  \psi(x) \partial_x \psi(x)  \,.
\ee
The normal ordering for singular products of operators can be written using a point-splitting regularization as   
\be
: \psi(x) \partial \psi(x): = \lim_{y \rightarrow x } (\psi(x) \partial \psi(y) - \expval{\psi(x) \partial \psi(y)}{\Omega})\,.
\ee
To use a more compact notation, in what follows we define 
\begin{equation}
    \widetilde{T}(x, y) = \frac{i}{2}  \psi(x) \partial_{y} \psi(y) \,.
\label{eq:def_T00(x,y)}
\end{equation}
Then, the normal ordering of the stress tensor is given by
\begin{equation}
    :T(x): = \lim_{y \rightarrow x } \widetilde{T}(x, y) - \expval{\widetilde{T}(x, y)}{\Omega} \, ,
\label{eq:T00_normalmenteordenado}
\end{equation}
and using Eq. (\ref{eq:correlador}) we have
\begin{equation}
    \expval{\widetilde{T}(x, y)}{\Omega} = \frac{1}{4 \pi (x - y)^{2}} \ \text{.}
\label{eq:correladorT}
\end{equation}
The generator of the twist (\ref{eq:deftwist2}) writes
\begin{equation}
    \widetilde{H} = \int_{\mathbb{R}} dx \ W^{\dagger} \left( :T(x): \otimes 1 \right) W\,.
\end{equation}
Using  (\ref{eq:T00_normalmenteordenado}) we get
\begin{equation}
\begin{split}
    \widetilde{H} &= \int_{\mathbb{R}} dx \lim_{y \rightarrow x} \ W^{\dagger} \left( \left( \widetilde{T}(x, y) - \expval{\widetilde{T}(x, y)}{\Omega} \right) \otimes 1 \right) W \\
    &= \int_{\mathbb{R}} dx \lim_{y \rightarrow x} \ W^{\dagger} \left( \widetilde{T}(x, y) \otimes 1 \right) W - \expval{\widetilde{T}(x, y)}{\Omega}\,.\label{integral}
\end{split}
\end{equation}
From this last expression we can see that there are several cases to consider. When $x, y \in A$, the operator $\widetilde{T}(x, y)$
is in the algebra  $\mathcal{A}(A)$. Then $W$ transforms it trivially to the same operator in the global Hilbert space, see (\ref{prop0}). Therefore we have 
\begin{equation}
    W^{\dagger} \left( \widetilde{T}(x, y) \otimes 1 \right) W = \widetilde{T}(x, y)\,, \qquad \ x,y \in A \,.
\label{eq:}
\end{equation}
When $x, y \notin A$ we use property (\ref{qq}), 
\begin{equation}
    W^{\dagger} \left( \widetilde{T}(x, y) \otimes 1 \right) W = J_{\mathcal{AB}} \ J_{\mathcal{A}} \ \widetilde{T}(x, y)  \ J_{\mathcal{A}} \ J_{\mathcal{AB}}\,,\qquad \ x,y \notin A \,.
\end{equation}
A last case is when  $x \in A$ but $y \notin A$ or vice versa. In this case, we have to express $\widetilde{T}(x, y)$ in terms of the fermionic operators. For example, suppose $x \in A$, $y \notin A$,
\begin{equation}
\begin{split}
    W^{\dagger} \left( \psi(x) \partial \psi(y) \otimes 1 \right) W &= W^{\dagger} \psi(x)  \otimes 1 W \ W^{\dagger} \partial \psi(y) \otimes 1 W \\ 
    &= \psi(x) \ J_{\mathcal{AB}} \ J_{\mathcal{A}} \ \partial \psi(y)  \ J_{\mathcal{A}} \ J_{\mathcal{AB}} \ \text{.}
\end{split}
\end{equation}
When $x \notin A$ but $y \in A$ we have a similar expression where $J_{\mathcal{AB}} \ J_{\mathcal{A}}$
now act on $\psi(x)$.
We will write
\begin{equation}
    W^{\dagger} \left( \widetilde{T}(x, y) \otimes 1 \right) W =  \widetilde{T}_{1,J}(x, y) \,,\qquad \ x \in A \ \text{y} \ y \notin A \,,
\end{equation}
 where $\widetilde{T}_{1,J}(x, y)$ indicates that only the fermion evaluated at  $y$ transforms with $J_{\mathcal{AB}} \ J_{\mathcal{A}}$. 
 
Hence, to evaluate the integral (\ref{integral}) we have to take into account the three cases. It might seem that the third case, when one of the variables is inside $A$ and the other outside, is marginal and will not give contributions because ultimately we are going to set $y\rightarrow x$. However, we will see this part of the integral provides an important numerical contribution to the twist Hamiltonian. Of course, this constant is irrelevant when computing twist transformations of operators.   

To get a more symmetric expression let us rename the variables as $x \rightarrow x-h/2$ y $y \rightarrow x+h/2$. The point-splitting limit is $h \rightarrow 0$. Separating the integral in the different cases we have 
\begin{equation}
\begin{split}
    \widetilde{H} = &\lim_{h \rightarrow 0 } \int_{- a + \frac{h}{2}}^{{a - \frac{h}{2}}} \ \left( \widetilde{T}(x - \frac{h}{2}, x + \frac{h}{2}) - \expval{\widetilde{T}(x - \frac{h}{2}, x + \frac{h}{2})}{\Omega} \right) dx\\
    &+ \int_{a + \frac{h}{2}}^{\infty} \ \left( J_{\mathcal{AB}} \ J_{\mathcal{A}} \ \widetilde{T}(x - \frac{h}{2}, x + \frac{h}{2}) \ J_{\mathcal{A}} \ J_{\mathcal{AB}} - \expval{\widetilde{T}(x - \frac{h}{2}, x + \frac{h}{2})}{\Omega} \right) dx \\
    &+ \int_{-\infty}^{-a - \frac{h}{2}} \ \left( J_{\mathcal{AB}} \ J_{\mathcal{A}} \ \widetilde{T}(x - \frac{h}{2}, x + \frac{h}{2}) \ J_{\mathcal{A}} \ J_{\mathcal{AB}} - \expval{\widetilde{T}(x - \frac{h}{2}, x + \frac{h}{2})}{\Omega} \right) dx \\
    &+ \int_{a - \frac{h}{2}}^{{a + \frac{h}{2}}} \ \left( \widetilde{T}_{1,J}( x - \frac{h}{2} , x + \frac{h}{2})  - \expval{\widetilde{T}(x - \frac{h}{2}, x + \frac{h}{2})}{\Omega} \right) dx \\
    &+ \int_{- a - \frac{h}{2}}^{{-a + \frac{h}{2}}} \ \left( \widetilde{T}_{J,1}(x - \frac{h}{2} , x + \frac{h}{2} )  - \expval{\widetilde{T}(x - \frac{h}{2}, x + \frac{h}{2})}{\Omega} \right) dx \,.
\end{split}
\label{eq:generador_twist_integral}
\end{equation}
The limit of the first term is simply 
\begin{equation}
    \lim_{h \rightarrow 0 } \int_{- a + \frac{h}{2}}^{{a - \frac{h}{2}}} \ \left( \widetilde{T}(x - \frac{h}{2}, x + \frac{h}{2}) - \expval{\widetilde{T}(x - \frac{h}{2}, x + \frac{h}{2})}{\Omega} \right) dx = \int_{A} :T(x): dx \ \text{.}
\label{eq:limite_primer_termino}
\end{equation}
In the second and third terms we find the transformations of the full operator, but since $J_{\mathcal{AB}}$ and $J_{\mathcal{A}}$
leave the vacuum invariant, we get
\begin{equation}
    \expval{\widetilde{T}(x - \frac{h}{2}, x + \frac{h}{2})}{\Omega} = \expval{J_{\mathcal{AB}} \ J_{\mathcal{A}} \ \widetilde{T}(x - \frac{h}{2}, x + \frac{h}{2}) \ J_{\mathcal{A}} \ J_{\mathcal{AB}}}{\Omega}
    \,.
\end{equation}
Therefore the second term gives
\begin{equation}
\begin{split}
    &\lim_{h \rightarrow 0} \left[ \int_{a + \frac{h}{2}}^{\infty} \ \left( J_{\mathcal{AB}} \ J_{\mathcal{A}} \ \widetilde{T}(x - \frac{h}{2}, x + \frac{h}{2}) \ J_{\mathcal{A}} \ J_{\mathcal{AB}} - \expval{\widetilde{T}(x - \frac{h}{2}, x + \frac{h}{2})}{\Omega} \right) dx \right. \\
    &\left. + \int_{-\infty}^{-a - \frac{h}{2}} \ \left( J_{\mathcal{AB}} \ J_{\mathcal{A}} \ \widetilde{T}(x - \frac{h}{2}, x + \frac{h}{2}) \ J_{\mathcal{A}} \ J_{\mathcal{AB}} - \expval{\widetilde{T}(x - \frac{h}{2}, x + \frac{h}{2})}{\Omega} \right) dx \right]\\
    &= \int_{A^{\prime}} \ :J_{\mathcal{AB}} \ J_{\mathcal{A}} \ T(x) \ J_{\mathcal{A}} \ J_{\mathcal{AB}}: dx
    \,.
\end{split}
\end{equation}
Then we can take normal ordering after computing the transformation of $T(x)$. Recalling (\ref{eq:twist_estandar_1}) we have

\begin{eqnarray}
  &&  J_{\mathcal{AB}} \ J_{\mathcal{A}} \psi(x) J_{\mathcal{A}} \ J_{\mathcal{AB}} = c_{+}(x) \psi(q_{+}(x)) + c_{-}(x) \psi(q_{-}(x))\,,
\label{eq:aplicacion_Js}
\\ && q_{\pm}(x) = \frac{x \left( - a^{2} (b-a)^{2} \pm x \sqrt{4 a b (a b - \frac{a^{4}}{x^{2}})^{2} + (b-a)^{4} \frac{a^4}{x^{2}}}\right)}{2 (a b x^{2} - a^4)}\,,
\\ && c_{\pm}(x) = \frac{a}{x} \frac{1}{q_{\pm}(x)-\frac{a^2}{x}} \left(\frac{b}{b^{2} - q_{\pm}(x)^{2}} + \frac{a}{q_{\pm}(x)^{2} - a^2} \right)^{-1} \ \text{.}
\end{eqnarray}
We recall that $q_\pm$ maps the complement $A'$ of $A$ to the intervals $Z_\pm$. The functions $c_\pm$ have domain $A'$ and $|c_\pm|\le 1$.  
From here, using (\ref{eq:def_T00(x,y)}), we arrive at
\begin{equation}
\begin{split}
    :\widetilde{J}_{\mathcal{AB}} \ \widetilde{J}_{\mathcal{A}} T(x) \widetilde{J}_{\mathcal{A}} \ \widetilde{J}_{\mathcal{AB}}:
    = &\sum_{i = \{+ , - \}} c_{i}(x)^{2} q_{i}^{\prime}(x) \, :T(q_{i}(x)): \\
    & + \frac{i}{2} \sum_{i,j \in \{+ , - \}, \, i \neq j} :c_{i}(x) \psi(q_{i}(x)) \left( c_{j}(x) \psi(q_{j}(x)) \right)^{\prime}:
    \ \text{.}
\end{split}
\label{eq:JT00J_final1}
\end{equation}
The full derivation of this equation can be consulted in appendix  \ref{A:twist_generator_desarrollo_T00}. The first term is again proportional to the stress tensor while the second one is a non-local product of two fermion fields. We now make more explicit the expressions of these terms. 

\begin{figure}[H] 
     \centering 
     \makebox[\textwidth][c]{\includegraphics[width=0.5\textwidth]{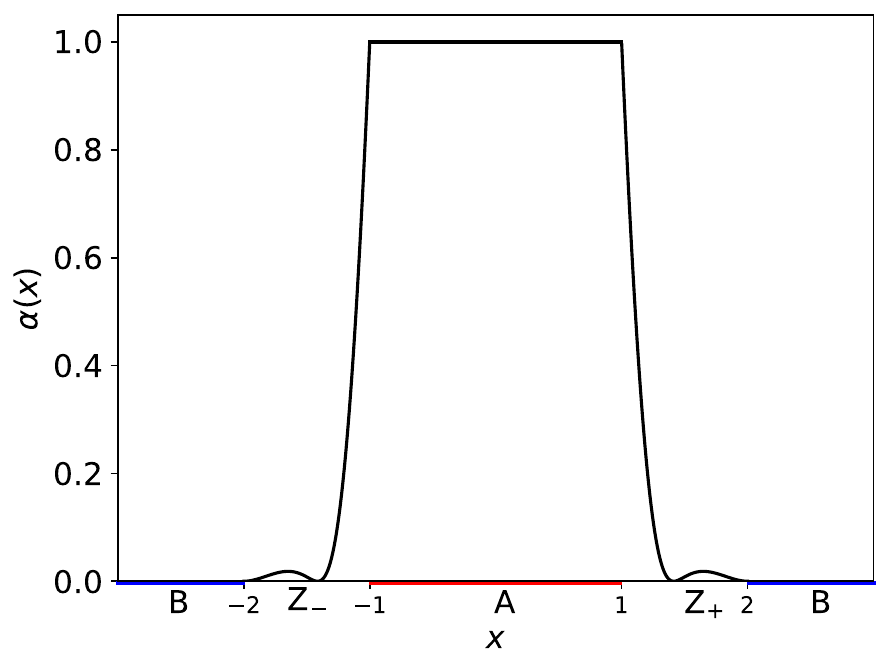}} 
     \caption{Function $\alpha(x)$ in eq.  (\ref{eq:def_f_soporteacotado})
     giving the smearing factor of the energy density in the local term of the twist generator. Here we have chosen the intervals as $A =(-1, 1)$, $Z_{-} = (-2, -1)$, $Z_{+} = (1, 2)$, $B = (-\infty, -2) \cup (2, \infty)$.
     } 
     \label{fig:f_soporteacotado} 
\end{figure}

If $x \notin A$ then $q_{\pm}(x) \in Z_{\pm}$. Hence the support of the terms linear in the stress tensor includes $A$, eq. (\ref{eq:limite_primer_termino}),  and the smearing zones. This is made more explicit by the change of variables  $q = q_{i}(x)$ in the first term of (\ref{eq:JT00J_final1}):
\begin{equation}
    \sum_{i \in \{+ , -\}} \int_{A^{\prime}} \ dx  \ c_{i}^{2}(x) q_{i}^{\prime}(x) :T(q_{i}(x)):
    = \sum_{i \in \{+ , -\}} \int_{Z_{i}} \ dq \ c_{i}^{2}(q_{i}^{-1}(q)) :T(q): \,,
\end{equation}
where $q_i^{-1}$ is the inverse function of $q_i$.
Adding this term with (\ref{eq:limite_primer_termino}) we get the local part of the twist generator,
\begin{equation}
 \widetilde{H}_{\textrm{loc}} = \int_{A \cup Z} \alpha(x) :T(x): dx \,,
\label{eq:integral_T_soporteacotado_1}
\end{equation}
where
\begin{equation}
    \alpha(x) = 
    \begin{cases}
        \quad 1 &  \, \qquad x \in A \\
        g_{\pm}(x)^{2} &  \, \qquad x \in Z_{\pm} \\
    \end{cases}
    \,,
\label{eq:def_f_soporteacotado}
\end{equation}
and
\begin{equation}
  g_\pm(x)=  c_{\pm}(q_{\pm}^{-1}(x)) = \frac{2 a^2 b^2- (a^2 + b^2) x^2 \pm x \sqrt{(b - a)^4 x^2+ 4 a b (x^2- a b)^2}}{2 a (b - a	) (a b +x^2)}\,.
 \label{101}
\end{equation}
Then $\alpha(x)=\alpha(-x)$ is even and $g_-(x)=g_+(-x)$. An example of the function $\alpha(x)$ is shown in the figure \ref{fig:f_soporteacotado}. Remarkably, this smearing function has discontinuous first derivatives at the end-points $\pm a$ of the interval $A$.  
The standard twist then differs from the twists (\ref{eq:integral_T_soporteacotado_1}) formed by simply smearing $T$ with smooth functions by this feature, on top of having additional non-local terms in the buffer zone. 

Then, for simplifying the non-local term (the last term in (\ref{eq:JT00J_final1})), we can integrate by parts one of the terms of the sum,
anti-commute the fermions to get a unique term and then change variables to obtain an integral over the buffer zone. We get
\bea
 && \hspace{-.4cm}\tilde{H}_{\textrm{no loc}}= \frac{i}{2} \, \sum_{i,j \in \{+ , -\}, \  i \neq j} \int_{A^{\prime}} dx \, :c_{i}(x) \psi(q_{i}(x)) \left(c_{j}(x) \psi(q_{j}(x)) \right)^{\prime}: \, \nonumber\\
 && \hspace{-.7cm} = \frac{i}{2} \, \int_{A^{\prime}} dx \, :c_{-}(x) \psi(q_{-}(x)) \left(c_{+}(x) \psi(q_{+}(x)) \right)^{\prime} + c_{+}(x) \psi(q_{+}(x)) \left(c_{-}(x) \psi(q_{-}(x)) \right)^{\prime}: \, \nonumber\\
 && \hspace{-.7cm} = i \, \int_{A^{\prime}} dx \, :c_{-}(x) \psi(q_{-}(x)) \left(c_{+}(x) \psi(q_{+}(x)) \right)^{\prime}: \, \nonumber\\
 && \hspace{-.7cm} = i \, \int_{Z_{+}} dx \, :c_{-}(q^{-1}_{+}(x)) \psi(q_{-}(q^{-1}_{+}(x))) \left(c_{+}(q^{-1}_{+}(x)) \psi(x) \right)^{\prime}: \, \nonumber\\
 && \hspace{-.7cm} = i \, \int_{Z_{+}} dx \, :g_{-}(\tilde{x}) \psi(\tilde{x}) \left(g_{+}(x) \psi(x) \right)^{\prime}:=-i \, \int_{Z_{-}} dx \, :\left(g_{-}(x) \psi(x)\right)' g_{+}(\tilde{x}) \psi(\tilde{x}): \, .
\eea
In the last equation we have written  
\be
\tilde{x}=q_-(q_+^{-1}(x))=- ab/x= q_+(q_-^{-1}(x))\,,\qquad \tilde{\tilde{x}}=x\,.
\ee
The function $\tilde{x}$ maps $Z_{+}\leftrightarrow Z_{-}$. 
The non-local term consists of a bilinear of the fermion fields with one field in each of the intervals $Z_{+}, Z_{-}$. 

It rests to evaluate the last two terms in (\ref{eq:generador_twist_integral}), which are localized in the boundary of the interval. Let us look how  $\widetilde{T}(x, y)$ transforms in this case. Let us first see what happens for the term localized near the first boundary in $x=-a$. We have
\begin{equation}
    \widetilde{T}_{J,1}(x - \frac{h}{2}, x + \frac{h}{2}) = \sum_{i \in \{+ , -\}} c_{i}(x - \frac{h}{2}) \widetilde{T}(q_{i}(x - \frac{h}{2}), x + \frac{h}{2}) \ \text{.}
\end{equation}
The integral gives
\begin{equation}
    \int_{-a - \frac{h}{2}}^{-a + \frac{h}{2}} \left(\sum_{i \in \{+ , -\}} c_{i}(x - \frac{h}{2}) \widetilde{T}(q_{i}(x - \frac{h}{2}), x + \frac{h}{2}) - \expval{\widetilde{T}(x - \frac{h}{2}, x + \frac{h}{2})}{\Omega}\right) dx\ \text{.}
\end{equation}
We have to analyze this integral with some care. Even if the size of the interval vanishes with $h\rightarrow 0$,
it turns out there are divergent contributions of the integrand in this limit, giving place to a non-trivial 
constant contribution. We normal order the operator extracting its expectation value, using eq.(\ref{eq:correladorT}) we have
\begin{equation}
    \expval{\widetilde{T}_{J,1}(x - \frac{h}{2}, x + \frac{h}{2})}{\Omega} = \sum_{i \in \{+ , -\}}  \frac{ c_{i}(x - \frac{h}{2})}{4 \pi (q_{i}(x - \frac{h}{2}) - (x + \frac{h}{2}))^2} \ \text{.}
\end{equation}
Adding and subtracting this term gives an integral of a normal order operator in a vanishing small interval, plus a constant. Taking the limit $h\rightarrow 0$ eliminates the operator and leaves the contribution 
\begin{equation}
   \lim_{h\rightarrow 0}\, \int_{-a - \frac{h}{2}}^{-a + \frac{h}{2}}  \left(-\frac{1}{4 \pi h^{2}} + \sum_{i \in \{+ , -\}}  \frac{c_{i}(x - \frac{h}{2})}{4 \pi (q_{i}(x - \frac{h}{2}) - (x + \frac{h}{2}))^2} \right) dx\ \text{.}
\end{equation}
Here we have used $\expval{T(x - \frac{h}{2}, x + \frac{h}{2})}{\Omega} = \frac{1}{4 \pi h^{2}}$. 

The next simplification is that the term with  $q_{+}(x)$ does not have divergences for $h\rightarrow 0$ and hence does not contribute when the integration interval collapses to a point. This is because the image of this function belongs to  $Z_{+}$, and the denominator is bounded above some constant value. We write
\begin{equation}
    f_{-}(x, h) = \frac{-1}{4 \pi h^{2}} + \frac{ c_{-}(x - \frac{h}{2})}{4 \pi ((q_{-}(x - \frac{h}{2}) - (x + \frac{h}{2})))^2} \,,
\end{equation}
and 
\begin{equation}
    \int_{ - a - \frac{h}{2}}^{ - a + \frac{h}{2}} f_{-}(x, h) dx = \int_{- \frac{h}{2}}^{\frac{h}{2}} f_{-}(-a + u , h) du \ \text{.}
\end{equation}
Expanding for small values of $u$ we have 
\begin{equation}
\begin{split}
    \int_{- \frac{h}{2}}^{\frac{h}{2}} f_{-}( - a + u , h) du
    &= \int_{- \frac{h}{2}}^{\frac{h}{2}} \left[f_{-}( - a, h) + f_{-}^{\prime}( - a, h) u + f_{-}^{\prime \prime}(-a, h) \frac{u^{2}}{2!} + O(u^{3})\right] du \\
    &= f_{-}( - a, h) h + f_{-}^{\prime \prime}(-a, h) \frac{h^{3}}{24} + \cdots
    \,.
\end{split}
\label{eq:serie_integral_fm}
\end{equation}
To evaluate these terms we use that
\begin{equation}
    q_{\pm}(\pm a) = \pm a, \quad q_{\pm}^{\prime}(\pm a) = 1, \quad c_{\pm}(\pm a) = 1, \quad c_{\mp}(\pm a) = 0 , \quad c^{\prime}_{\pm}(\pm a) = \frac{q_{\pm}^{\prime \prime}(\pm a)}{2} \ \text{.}
\label{eq:propiedades_qi}
\end{equation}
 It is only necessary to compute the first two terms in the expansion. It can be shown that the terms of larger order vanish in the limit (see appendix  \ref{A:limitesfpm}).
 We have
\begin{equation}
\begin{split}
    &\lim_{h \rightarrow 0} f_{-}(-a, h) h = 0 \,,\\
    &\lim_{h \rightarrow 0} f_{-}^{\prime \prime}(-a, h) \frac{h^{3}}{24} = \frac{q_{-}^{\prime \prime}(-a)}{48 \pi} = \frac{b}{6 \pi (b^{2} - a^{2})} \,.
\end{split}
\label{eq:limites_fm}
\end{equation}
It then follows that
\begin{equation}
 \lim_{h\rightarrow 0}\,   \int_{ - a - \frac{h}{2}}^{ - a + \frac{h}{2}} f_{-}(x, h) dx = \frac{b}{6 \pi (b^{2} - a^{2})} \ \text{.}
\end{equation}

Now we have to do a similar calculation for the integral near the boundary  $x=a$. In this case, the field that falls out of the region $A$ is a field derivative and taking into account that only the term with $q_{+}(x)$ in the denominator has a divergence, we get the integrand
\begin{equation}
    f_{+}(x, h) = -\frac{1}{4 \pi h^{2}} + \frac{c_{+}^{\prime}(x + \frac{h}{2})}{4 \pi (x - \frac{h}{2} - q_{+}(x + \frac{h}{2}))}
    + \frac{c_{+}(x + \frac{h}{2}) q_{+}^{\prime}(x + \frac{h}{2})}{4 \pi (x - \frac{h}{2} - q_{+}(x + \frac{h}{2}))^{2}}  \ \text{.}
\end{equation}
Following the same ideas as above, the expansion gives
\begin{equation}
    \begin{split}
        &\lim_{h \rightarrow 0} f_{+}(a, h) h = 0 \,,\\
        &\lim_{h \rightarrow 0} f_{+}^{\prime \prime}(a, h) \frac{h^{3}}{24} = - \frac{q_{+}^{\prime \prime}(a)}{48 \pi} = \frac{b}{6 \pi (b^{2} - a^{2})} \,.
    \end{split}
\label{eq:limites_fp}
\end{equation}
As a result, the integral is 
\begin{equation}
  \lim_{h\rightarrow 0}\,  \int_{a - \frac{h}{2}}^{a + \frac{h}{2}} f_{+}(x, h) dx = \frac{b}{6 \pi (b^{2} - a^{2})} \ \text{.}
\end{equation}
Then the contribution of the two boundaries is the same and the full constant terms is twice this value.   

Summarizing, the expression of the twist generator is
\begin{equation}
    \widetilde{H} = \frac{1}{3 \pi} \frac{b}{b^{2} - a^{2}} + \int_{A \cup Z} dx \ \alpha(x) :T(x): 
    + \,i \,  \int_{Z_+} dx \,   :g_{-}(\tilde{x})\psi(\tilde{x})\,\, (g_{+}(x)\psi(x))^{\prime}: \,,
\label{tg}
\end{equation}
where $\alpha(x)$ and $g_\pm(x)$ are given in (\ref{eq:def_f_soporteacotado}), (\ref{101}).

This twist generator has an interesting property \cite{buchholz1986noether}. As it results from a unitary transformation from the ordinary Hamiltonian $H$ which is positive definite, it is also positive definite. In fact, it has the same spectrum as the original Hamiltonian even if it has support in a finite interval.\footnote{In this case the spectrum is $\mathbb{R}^+$. Interestingly, the analogous single interval twist generator inside the circle must have discrete spectrum $\mathbb{Z}+$const.,  as the corresponding circle Hamiltonian. } In particular, the vacuum expectation value must be positive, which is immediate from the normal ordering of the operators in (\ref{tg}). We have
\begin{equation}
    \expval{\widetilde{H}}{\Omega} = \frac{1}{3 \pi} \frac{b}{b^{2} - a^{2}}
     \ \text{.}
\end{equation}
It is interesting to note that using (\ref{eq:def_f_soporteacotado}), (\ref{eq:propiedades_qi}), (\ref{eq:limites_fm}) y (\ref{eq:limites_fp})
it follows that this constant value can also be written in terms of the change of the slope of the smearing function of the local term on the points where it becomes non-differentiable:
\begin{equation}
    \expval{\widetilde{H}}{\Omega} = \frac{\alpha^{\prime}_{-}(-a) - \alpha^{\prime}_{+}(a)}{48 \pi}
     \ \text{,}
\end{equation}
where $\alpha'_+$ means the limit on the right of the derivative of the smearing function $\alpha(x)$ and $\alpha'_-$ the limit on the left. 

\subsection{Energy inequality and comparison with Fewster-Hollands bound}

It is known that, for any QFT, smearing the energy density with a positive function of compact support such as
\be
\int dx \, \alpha(x)\, T_{00}(x)\,,
\ee
produces an operator that cannot be positive. In other words, there are always states such that this localized energy turns out to be negative. The reason is quite simple. If this operator were positive definite, since its vacuum expectation value vanishes, it would be the case that the operator annihilates the vacuum. But it is not possible to annihilate the vacuum for a localized operator because of the Reeh-Schlieder theorem (for a simple account see \cite{witten2018aps}). 

It has been of interest to understand how much localized negative energy can a state have. This quest gave place to the quantum energy inequalities, that generically give a (negative) lower bound on the localized energy in terms of the smearing function $\alpha$ (for a review see \cite{fewster2012lectures}). The most general and sharpest of such bounds was obtained by Fewster and Hollands (FH bound) for CFT's in two dimensions \cite{fewster2005quantum}. It reads
\be
\int dx \, \alpha(x)\, \langle\Phi| T(x)|\Phi\rangle\ge -\frac{c}{12 \pi}\, \int dx\, \left(\frac{d \sqrt{\alpha(x)}}{dx}\right)^2\,,  
\ee
for any vector state $|\Phi\rangle$, and  where $c$ is the central charge. For the free Majorana fermion $c=\frac{1}{2}$. The bound follows from the positivity of the Hamiltonian unitarily transformed by a general conformal transformation (diffeomorphism of the line). The term on the right hand side is produced by the anomaly. The bound is sharp because it is saturated for the corresponding unitary transformation of the vacuum state. 

Different types of bounds on the energy density, involving entropy quantities, have also been explored. For impure global states, sharper bounds containing an entropy term have been derived \cite{blanco2018modular}. Another type of bound is the quantum null energy condition, which bounds the energy density at a point in terms of a second derivative of an entanglement entropy \cite{bousso2016quantum}.   

From the standard translation twists it follows a different type of energy bounds that we can call an operator-bounded energy inequality. These involve the energy density in a bounded region of space and some specific operator, that is not given in terms of the energy density, at the boundary of this region, or more precisely in the smearing zone. The derivation of the bound follows a similar idea as in the FH bound. It starts from the positivity of the Hamiltonian and a unitary transformation that preserves the spectrum. In this case, however, the transformation is the localization transformation $W$. This maps the Hamiltonian on one copy of a duplicated Hilbert space to a twist. This twist essentially contains the energy density in the localization region plus some new elements at the boundary. 

For the free chiral fermion we can be more specific. Given the positivity of the twist generator spectrum, for any state $|\Phi\rangle$ we have, using the single interval twist of the previous section,  
\begin{equation}
\begin{split}
    &\int_{A \cup Z} dx \ \alpha(x)  \expval{:T(x):}{\Phi} \ge \\
    &\hspace{2cm}- \frac{1}{3 \pi} \frac{b}{b^{2} - a^{2}}
    -   \,i \,  \int_{Z_{+}} dx \,   \langle \Phi|:g_{-}(\tilde{x})\psi(\tilde{x}) \,\, (g_{+}(x)\psi(x))^{\prime}:|\Phi\rangle\,,
\end{split}
\label{eq:condicionenergia}
\end{equation}
where $\alpha(x)\in [0,1]$ is given by eq. (\ref{eq:def_f_soporteacotado}), and $g_\pm(x)$, with $|g_\pm(x)|\le 1$,  in eq. (\ref{101}). 

This bound is also sharp because it follows from a unitary transformation of the Hamiltonian (in a duplicated Hilbert space). Indeed, it will be saturated by any state that is the unitary transformation of the vacuum state in the first Hilbert space. This is any state of the form
\be
W^\dagger \, (|\Omega\rangle\otimes |\Psi\rangle)=|\Omega\rangle_{\cal N}\otimes |\Psi\rangle_{{\cal N}'} \,, \label{unen}  
\ee   
for any $|\Psi\rangle$. These types of states have the special feature that they are unentangled between the algebras of ${\cal A}(A)\subset {\cal N}$ and ${\cal A}(B)\subset {\cal N}'$. More precisely, these states do not have any connected correlation between $A$ and $B$.  

In this sense, this bound and FH bound, when applied to the same smearing function $\alpha$ of eq. (\ref{eq:def_f_soporteacotado}), are not comparable. For an unentangled state of the form (\ref{unen}) the bound (\ref{eq:condicionenergia}) saturates while the FH does not. Conversely, for a limit of states obtained from conformal transformations of the vacuum such that the FH bound saturates, the bound (\ref{eq:condicionenergia}) does not. 

\begin{figure}[t] 
     \centering 
     \makebox[\textwidth][c]{\includegraphics[width=0.5\textwidth]{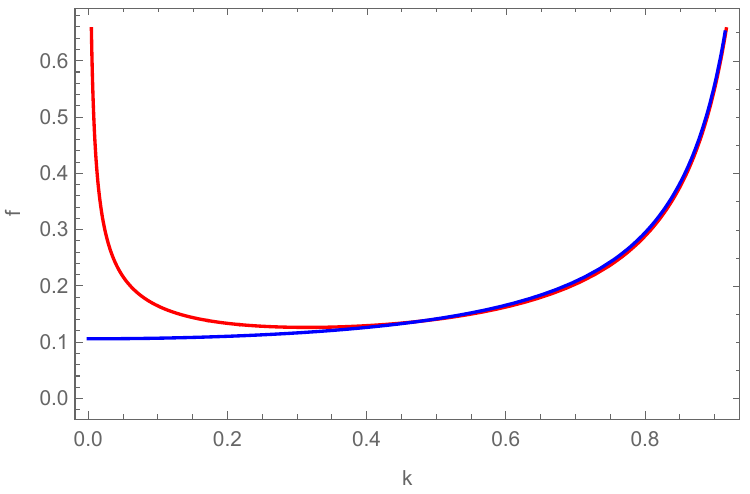}} 
     \caption{Functions $f_{FH}(k)$ (red) and $f(k)$ (blue). The difference between the functions changes sign around $k\sim 0.47$. In the limit $k\rightarrow 1$ both functions have the same pole divergence but differ by a constant. } 
     \label{figfin} 
\end{figure}

For our present smearing function $\alpha$, the constant on the left hand side of the FH bound is finite and can be computed exactly, though we refrain from writing this long expression here.\footnote{When $\alpha=0$ the integrand in the FH bound has to be taken zero. See \cite{fewster2005quantum} for mathematical details. In particular, in that paper, $\alpha(x)$ is assumed to be a smooth function of fast decrease. 
Our smearing function is not smooth, but we have taken the view that the FH holds as a limit whenever the right hand side remains bounded.} To compare with the constant term in the left hand side of (\ref{eq:condicionenergia}) we can express both constant terms in dimensionless form writing $a= k\, b$, $0<k<1$. Then we express
\bea
\frac{1}{24 \pi}\, \int dx\, \left(\frac{d \sqrt{\alpha(x)}}{dx}\right)^2=b^{-1}\, f_{FH}(k)\,,
 \eea
and analogously for the constant term in our inequality  
\be
\expval{\widetilde{H}}{\Omega} =b^{-1}\, f(k)\,,\qquad f(k)=\frac{1}{3\pi}\, \frac{1}{1-k^2}\,.
\ee
Figure \ref{figfin} shows the two functions. In the limit of a short buffer zone, $k\rightarrow 1$, both functions diverge in the same way while the difference converges to a constant,
\bea
f_{FH}(k) &\sim & (6 \pi)^{-1}\,(1-k)^{-1}+(8 \pi)^{-1}-1/48\,,\\
f(k) &\sim & (6 \pi)^{-1}\,(1-k)^{-1}+(12 \pi)^{-1}\,,
\eea
such that  $f-f_{FH}\rightarrow 0.00757\cdots$.  In the opposite limit, $k\rightarrow 0$, the function  $f_{FH}\sim (24 \sqrt{k})^{-1}$ diverges, while $f\sim 1/(3\pi)$ remains bounded. 

It is interesting to note that these operator bounds can be greatly generalized by combining the transformations. First one makes a conformal transformation of the Hamiltonian in the first factor Hilbert space and then applies the localization map. The result is a positive operator that has support in $A\cup Z$ and has a quite arbitrary smearing function for the energy density inside $A$. It also contains a bilinear of the fermion fields in the buffer zone. Another possibility is to use the localization map repeatedly transforming positive operators into positive operators. We will not pursue these constructions more explicitly here.

\section{Conclusions} 

We have computed explicitly the standard translation twists for the chiral fermion theory. These twists have interesting properties. One can continuously translate an operator from an interval arriving finally to a disjoint one without ever having passed through a gap between them. They also have a positive generator. This can be used to write operator-bounded energy inequalities. These properties are not shared by twists constructed by simply smearing the stress tensor. 

To generalize this construction to other theories is necessary to know the action of the modular conjugation for more than one interval. The explicit form of the modular Hamiltonian for two intervals is known for the chiral scalar too \cite{arias2018entropy}, and it would not be difficult to understand the action of the modular conjugation in this case. This action will also be linear in the fields, and as a result, the twist generator will be quadratic. However, it is expected that this action and the resulting twist generators would be completely non-local in the buffer zone. 

The chiral scalar algebras are subalgebras of the Dirac (complex) chiral fermion theory due to bosonization. This raises an interesting question. Twists for the scalar can be used in the fermionic theory. They will translate correctly the current operator and the stress tensor, which are shared by the fermion model. It is interesting to understand what would be the action on the fermions fields. What is clear is that at least these twists for the scalar will be unable to move fermions between two disjoint intervals.\footnote{That is, the twist formed by a split between the additive algebra for the intervals and its commutant algebra separated by a buffer zone.} This is because the operator content of the twists belongs to the algebras generated by the two intervals and their buffer zones and cannot contain a fermion operator in one interval and an antifermion one on the other. However, this is necessary for translating the fermion operator. More generally, a twist that produces jumps between two disjoint regions always has to contain all types of charge-anticharge pairs for the charges that are able to transport.  
In consequence, the scalar twists will translate the fermions to the buffer zone but will be unable to produce the jump, effectively filtering the charged part of the operator. It would be interesting to explore this more explicitly.

\section*{Acknowledgements}
This work was partially supported by CONICET, CNEA and Universidad Nacional de Cuyo, Argentina.

\appendix

\section{Modular conjugation from modular flow}
\label{mcmf}

Let us  write 
\begin{equation}
    \widetilde{\psi}(y) = \widetilde{J} \psi(y) \widetilde{J} =\int dx \ K(x, y)\, \psi(x) \,.\label{67}
\end{equation}
The anticommutator selects the kernel $K$:
\begin{equation}
    \{\psi(x), \widetilde{\psi}(y) \} = K(x, y) \,.\label{68}
\end{equation}
Using that $\{ \psi(x), \widetilde{\psi}(y) \} = 
\expval{\{ \psi(x), \widetilde{\psi}(y) \}}{\Omega}$, we get
\begin{equation}
    \expval{\{ \psi(x), \widetilde{\psi}(y) \}}{\Omega} = 
    \expval{\psi(x)\widetilde{J}\psi(y)\widetilde{J}}{\Omega} + \expval{\widetilde{J}\psi(y)\widetilde{J}\psi(x)}{\Omega} \ \text{.}
\end{equation}
Using that $\tilde{J}$ is antiunitary and that $\widetilde{J} \ket{\Omega} = \ket{\Omega}$, we have
\begin{equation}
    \{ \psi(x), \widetilde{\psi}(y) \} =     
        \expval{\psi(x)\widetilde{J}\psi(y)}{\Omega} +
        \expval{\psi(y)\widetilde{J}\psi(x)}{\Omega}^{*}
      \ \text{.}
\end{equation}
From this we get
\begin{equation}
\begin{split}
    \expval{\psi(x) \widetilde{J} \psi(y)}{\Omega}
    &= \expval{\psi(x) ZJ \psi(y)}{\Omega}
    = \expval{\psi(x) JZ^{\dagger} \psi(y)}{\Omega} \\
    &= \expval{\psi(x) J Z^{\dagger}\psi(y)Z Z^{\dagger}}{\Omega}
    = \expval{\psi(x)J i\Gamma \psi(y)}{\Omega} \\
    &= \expval{\psi(x)J i\Gamma \psi(y)\Gamma \Gamma}{\Omega}
    = \expval{\psi(x) J (-i) \psi(y)}{\Omega} \\
    &= i \expval{\psi(x) J \psi(y)}{\Omega}  \,,
\end{split}
\end{equation}
\begin{equation}
\begin{split}
    \expval{\psi(y) \widetilde{J} \psi(x)}{\Omega}^{*} 
    &= \expval{\psi(y) JZ^{\dagger} \psi(x)}{\Omega}^{*}
    = \expval{\psi(y) ZJ \psi(x)}{\Omega}^{*} \\ 
    &= \expval{Z Z^{\dagger} \psi(y) Z J \psi(x)}{\Omega}^{*}
    = \expval{ i \Gamma \psi(y) J  \psi(x)}{\Omega}^{*} \\
    &= (-i) \expval{\psi(y)  J  \psi(x)}{\Omega}^{*}
    = (-i) \expval{\psi(x)  J  \psi(y)}{\Omega}^{*} \,.
\end{split}
\end{equation}
From this we obtain
\begin{equation}
\begin{split}
    \{ \psi(x), \widetilde{\psi}(y) \} &=  i  \left[\expval{\psi(x) J \psi(y)}{\Omega} - \expval{\psi(x) J \psi(y)}{\Omega}^{*}\right] \\
    &=- 2 \ \mathrm{Im}\left(\expval{\psi(x)J\psi(y)}{\Omega}\right)
    = -2 \ \mathrm{Im}\left(\expval{\psi(x)\Delta^{\frac{1}{2}}\psi(y)}{\Omega}\right) \ \text{.}
\end{split}
\end{equation}
At this point we find it useful to recall the formula for the modular evolved correlator found in \cite{hollands2021modular}
\begin{equation}
    \expval{\psi(x)\Delta^{it}\psi(y)}{\Omega} =
    \frac{1}{2 \pi i(x - y)} \ \frac{\Pi_{b}(x)\Pi_{a}(y) - \Pi_{b}(y)\Pi_{a}(x)}{e^{\pi t}\Pi_{b}(x)\Pi_{a}(y) - e^{-\pi t} \Pi_{b}(y)\Pi_{a}(x)}  \,,
\end{equation}
where the polynomials $\Pi_a(x), \Pi_b(x)$, are defined in (\ref{productos}).
For $x$ and $y$ in complementary regions this expression extends analytically for $-\frac{1}{2} < \mathrm{Im}(t) < \frac{1}{2}$. We need to take the limit to the boundary of the analyticity domain $t \rightarrow  -\frac{i}{2}$ where the function will be singular for $x,y$ such that $\Pi_{b}(x)\Pi_{a}(y) + \Pi_{b}(y)\Pi_{a}(x) = 0$. 
 Using that
\begin{equation}
    \lim_{y \to 0^{+}} \frac{1}{x - iy} = \frac{1}{x} + i \pi \delta(x) \,,
\end{equation}
we arrive at
\begin{equation}
     \lim_{t \rightarrow -\frac{i}{2}}\expval{\psi(x)\Delta^{it}\psi(y)}{\Omega} = 
    \frac{ -1}{2(x - y)} \
    \left[
        \frac{1}{\pi} G(x,y)^{-1} +
        i \,\delta(G(x, y))
    \right] \,,
\end{equation}
with
\begin{equation}
    G(x, y) =  \frac{\Pi_{b}(x)\Pi_{a}(y) + \Pi_{b}(y)\Pi_{a}(x)}{\Pi_{b}(x)\Pi_{a}(y) - \Pi_{b}(y)\Pi_{a}(x)} \,.
\end{equation}
Then, the explicit expression of the anticommutator is 
\begin{equation}
    \{ \psi(x), \widetilde{\psi}(y) \}
    = \frac{ 1 }{ (x - y)} \delta(G(x, y))\,.
\end{equation}
Using (\ref{67}) and (\ref{68}) we get the formula (\ref{eq:JpsiJ_formula1}) for modular reflected field $\widetilde{\psi}$ quoted in the main text.

\section{Twist Generator}
\label{A:twist_generator}

In this appendix we present some calculations that were not detailed in the main text of section \ref{charges} to allow for more clarity of the exposition. 

\subsection{Calculation of $:J_{\mathcal{AB}} \ J_{\mathcal{A}} T(x) J_{\mathcal{A}} \ J_{\mathcal{AB}}:$}
\label{A:twist_generator_desarrollo_T00}
Here we compute  
$:J_{\mathcal{AB}} \ J_{\mathcal{A}} T(x) J_{\mathcal{A}} \ J_{\mathcal{AB}}:$.
We find it convenient to use a more symmetric form of $T(x)$
\begin{equation}
    T(x) = \frac{i}{4} \left[\psi(x) \partial_{x} \psi(x) - \partial_{x} \psi(x) \, \psi(x) \right] \,.
\end{equation}
 We recall the transformation of the fermion field under the action of 
 $J_{\mathcal{A}} \ J_{\mathcal{AB}}$
\begin{equation}
    J_{\mathcal{AB}} \ J_{\mathcal{A}} \psi(x) J_{\mathcal{A}} \ J_{\mathcal{AB}}
    = \sum_{j = \{+ , -\}} c_{j}(x) \psi(q_{j}(x)) \ \text{.}
\end{equation}
The action of $J = J_{\mathcal{AB}} \ J_{\mathcal{A}}$ over $\psi(x) \partial_x \psi(x)$ is
\begin{equation}
    J \psi(x) \partial_x \psi(x) J^{\dagger}
    = J \psi(x) J^{\dagger} \ \partial_x (J \psi(x) J^{\dagger}) 
    = \sum_{j, i \in \{+ , - \}} c_{j}(x) \psi(q_{j}(x)) \ \left( c_{i}(x) \psi(q_{i}(x)) \right)^{\prime}\,.
\end{equation}
Then 
\begin{equation}
\begin{split}
    &J T(x) J^{\dagger} = 
    \frac{i}{4} \sum_{j, i \in \{+ , - \}} \left[c_{j}(x) \psi(q_{j}(x)) \left( c_{i}(x) \psi(q_{i}(x)) \right)^{\prime} - \left( c_{i}(x) \psi(q_{i}(x)) \right)^{\prime}  c_{j}(x) \psi(q_{j}(x)) \right] \\
    & = \sum_{i \in \{+ , - \}} c_{i}(x)^{2} q_{i}^{\prime}(x) \, \frac{i}{4} \left[\psi(q_{i}(x)) \partial \psi(q_{i}(x)) - \partial \psi(q_{i}(x)) \, \psi(q_{i}(x)) \right] \\
    & \quad + \frac{i}{4} \sum_{j, i \in \{+ , - \}, \, j \neq i} \left[c_{j}(x) \psi(q_{j}(x)) \left( c_{i}(x) \psi(q_{i}(x)) \right)^{\prime} - \left( c_{i}(x) \psi(q_{i}(x)) \right)^{\prime}  c_{j}(x) \psi(q_{j}(x)) \right] \\
    & = \sum_{i \in \{+ , - \}} c_{i}(x)^{2} q_{i}^{\prime}(x) \, T(q_{i}(x)) + \frac{i}{2} \sum_{j, i \in \{+ , - \}, \, j \neq i} c_{j}(x) \psi(q_{j}(x)) \left( c_{i}(x) \psi(q_{i}(x)) \right)^{\prime} \,.
\end{split}
\end{equation}
In the last equation we used that $q_{\pm}(x) \in Z_{\pm}$, and $q_{+}(x) \neq q_{-}(x) \, \forall x \in \mathbb{R}$.  Therefore  $\psi(q_{+}(x)) \psi(q_{-}(x)) = - \psi(q_{-}(x)) \psi(q_{+}(x))$.  
Summarizing, we have
\begin{equation}
\begin{split}
    :J_{\mathcal{AB}} \ J_{\mathcal{A}} T(x) J_{\mathcal{A}} \ J_{\mathcal{AB}}:
    = &\sum_{i \in \{+ , - \}} c_{i}(x)^{2} q_{i}^{\prime}(x) \, :T(q_{i}(x)): \\
    & + \frac{i}{2} \sum_{i, j \in \{+ , - \}, \, i \neq j} :c_{i}(x) \psi(q_{i}(x)) \left( c_{j}(x) \psi(q_{j}(x)) \right)^{\prime}: \,.
\end{split}
\end{equation}

\subsection{The limit $\lim_{h \rightarrow 0} f^{n}_{\pm} h^{n+1}$}
\label{A:limitesfpm}

In eq. (\ref{eq:limites_fm}) we neglected terms in the Taylor expansion of order greater than the second. We show here that these terms vanish in the $h\rightarrow 0$ limit. The functions have the following structure:
\begin{equation}
    f(x, h) = \frac{1}{g(x, h)^{2}}, \quad g(a, h) \propto h, \quad g^{\prime}(a, h) \propto h, \quad g^{(n)}(a, 0) \neq 0 \ \text{if} \ n \geq 2 \ \text{.}
\end{equation}
where we suppose that the divergence is in the point $x=a$. Then, if we derive with respect to $x$, in all the terms of $f^{(n)}(a, h)$ we get a factor with the form
\begin{equation}
    \frac{g^{\prime}(a, h)^{k - m}}{g(a, h)^{2 + k}} \propto \frac{1}{h^{2 + m}} \ \text{with} \ 0 \leq k, \ 0 \leq m \leq k, \ k+m \leq n \,,
\end{equation}
where  $k$ and $m$ are natural numbers that are given by the numbers of derivatives of the numerator and denominator respectively. There is a term for each possible combination of $k$ and $m$ that respects the constraints. 
For  $n$ even, that is the case of the terms arising from eq. (\ref{eq:serie_integral_fm}),
we can show that $m_{max} = \frac{n}{2}$ is the greatest possible value of  $m$. To see this, we take any pair $(k, m)$ such that $k + m =n$ and necessarily $m \leq k$. Then, if $m \neq k$, we can get a new pair with larger $m$ by increasing the value of $m$ and reducing $k$ to the same amount, such as to keep $k + m =n$ constant.
We can do this again $\frac{k-m}{2}$ times, until reaching $m = k$. This implies that $m_{max} = \frac{n}{2}$.
Therefore, 
\begin{equation}
    f^{n}(a, h) h^{n + 1} \propto h^{n - 1 - m_{max}} = h^{\frac{n - 2}{2} } \ \text{.}
\end{equation}
In consequence, if $n > 2$ all limits $h\rightarrow 0$ vanish. 

\bibliography{EE}{}
\bibliographystyle{utphys}

\end{document}